\documentclass[fleqn,usenatbib]{mnras}

\usepackage{newtxtext,newtxmath}

\usepackage[T1]{fontenc}

\DeclareRobustCommand{\VAN}[3]{#2}
\let\VANthebibliography\thebibliography
\def\thebibliography{\DeclareRobustCommand{\VAN}[3]{##3}\VANthebibliography}

\newcommand{\xmm}{{\it XMM--Newton}}
\newcommand{\swift}{{\it Swift}}

\usepackage{graphicx}	
\usepackage{amsmath}	
\usepackage{multicol}
\usepackage{orcidlink} 
\usepackage[noabbrev,capitalize]{cleveref} 
\usepackage{booktabs}
\usepackage{subcaption}

\title[Modelling power spectra of AGN]{Modelling variability power spectra of active galaxies from irregular time series }

\author[Lefkir et al.]{
Mehdy Lefkir\,\orcidlink{0000-0002-6972-2429}$^{1}$\thanks{E-mail: ml556@leicester.ac.uk},
Simon Vaughan\,\orcidlink{0000-0003-4808-092X}$^{1}$,
Daniela Huppenkothen\,\orcidlink{0000-0002-1169-7486}$^{2,3}$, Phil Uttley\,\orcidlink{0000-0001-9355-961X}$^{3}$, \newauthor Vysakh Anilkumar\,$^{4}$ 
\\
$^{1}$School of Physics and Astronomy, University of Leicester, Leicester, LE1 7RH, UK\\
$^{3}$SRON Netherlands Institute for Space Research, Niels Bohrweg 4, 2333 CA Leiden, The Netherlands\\
$^{2}$Anton Pannekoek Institute for Astronomy, University of Amsterdam, Science Park 904, NL-1098 XH Amsterdam, The Netherlands\\
$^{4}$Leiden Observatory, Leiden University, Postbus 9513, 2300 RA Leiden, The Netherlands}
\date{Accepted XXX. Received YYY; in original form ZZZ}

\pubyear{2025}

\begin{document}
\label{firstpage}
\pagerange{\pageref{firstpage}--\pageref{lastpage}}
\maketitle

\begin{abstract}
    A common feature of Active Galactic Nuclei (AGN) is their random variations in brightness across the whole emission spectrum, from radio to $\gamma$-rays. Studying the nature and origin of these fluctuations is critical to characterising the underlying variability process of the accretion flow that powers AGN. Random timing fluctuations are often studied with the power spectrum; this quantifies how the amplitude of variations is distributed over temporal frequencies. Red noise variability -- when the power spectrum increases smoothly towards low frequencies -- is ubiquitous in AGN. The commonly used Fourier analysis methods, have significant challenges when applied to arbitrarily sampled light curves of red noise variability. Several time-domain methods exist to infer the power spectral shape in the case of irregular sampling but they suffer from biases which can be difficult to mitigate, or are computationally expensive. In this paper, we demonstrate a method infer the shape of broad-band power spectra for irregular time series, using a Gaussian process regression method scalable to large datasets. The power spectrum is modelled as a power-law model with one or two bends with flexible slopes. The method is fully Bayesian and we demonstrate its utility using simulated light curves. Finally, Ark 564, a well-known variable Seyfert 1 galaxy, is used as a test case and we find consistent results with the literature using independent X-ray data from \textit{XMM--Newton} and \textit{Swift}. We provide publicly available, documented and tested implementations in Python and Julia.
\end{abstract}

\begin{keywords}
    methods: numerical -- methods: statistical -- galaxies: active -- galaxies: individual: Ark 564 -- X-rays: galaxies
\end{keywords}



\section{Introduction}

An Active Galactic Nuclei (AGN) is a bright region located at the centre of many galaxies, powered by accretion of gas on to a supermassive black hole \citep{2017A&ARv..25....2P}. Observations of the central engine of AGN will contain information about the physics of mass accretion around a supermassive black hole, but understanding how to interpret the observed patterns and use them to study the accretion flow and black hole remains an open challenge.
The black hole mass is typically in the range $10^6-10^9\,M_\odot$, and the accretion process can produce radiation from radio to X-rays as well as long-lasting relativistic jets \citep{1984ARA&A..22..471R}. In some cases, it can even generate powerful winds that affect the host galaxy evolution \citep{2012ARA&A..50..455F}. The central engine is usually too small to be resolved, the exceptions are the recent radio observations with the Event Horizon Telescope that resolve the vicinity around the black hole in M87 and in the Galactic Centre \citep{2019ApJ...875L...1E,2022ApJ...930L..12E}. In all other cases, information about the physics and geometry of the central engine must be extracted from the light received: its spectrum, variability through time, and its polarization (or a combination of these).

A feature of many AGN is their variability in brightness. In almost all cases the variations appear random, without predictable trends or cycles, with larger amplitude variations on longer time-scales. Although the variations are random, we typically assume that they are drawn from a process which is {\it weakly stationary}. By {\it stationary} we mean the statistical properties of the process remain constant over time. By {\it weakly stationary} we mean that the first two {\it moments} of the random process do not change with time: the mean and the autocovariance.
The Fourier transform of the autocovariance of a random process is known as the power spectrum, it contains the same information but for many problems it is simpler to work with than the autocovariance. Weak stationary implies a time-constant power spectrum, where the power spectrum determines the contribution to the total variance of the process from random fluctuations on different time-scales (inverse temporal frequencies).

AGN variability is often described as `red noise' \citep{1981MNRAS.194..987M,1993MNRAS.261..612P}, meaning that the power spectrum of the variations rises towards low temporal frequencies, approximately following a power-law form $f^{-\alpha}$, with power-law slope $\alpha>1$ (and typically $\alpha>2$) at high frequencies.
In order for the total variance of the process to remain finite, the integral of the power spectrum over all frequencies must converge; this requires that the power spectrum at low frequencies has a slope $\alpha<1$ \citep{1978ComAp...7..103P}. The frequency of this flattening defines a characteristic time-scale, presumably set by the size and other properties of the central engine, and can be used to compare different AGN. In most cases, measuring this time-scale requires long, months--years duration monitoring campaigns consisting of light curves, which are sparsely and often irregularly sampled. Accurate power spectral measurement then requires the use of methods that can account for the effects of this sampling.

In X-ray astronomy, the standard approach is to work in the Fourier domain by estimating the power spectrum using the periodogram \citep{2022arXiv220907954B,1989ASIC..262...27V}.
This is most naturally applied to regularly sample time series data; one way to handle the sparse and irregular sampling of real data is to use forward-modelling. Here, simulations of light curves drawn from a process with a given power spectral shape are resampled to match the observed sampling pattern (e.g. \citealt{2002MNRAS.332..231U}), and thereby include in the modelling process any distortions arising from the time sampling.
Using these methods, X-ray power spectral bends or breaks have been reported for many AGN \citep{1999ApJ...514..682E,2002MNRAS.332..231U,2003ApJ...593...96M}, showing that the bend time-scale scales linearly with black hole mass. This $t_\mathrm{bend} \propto M_{BH}$ relation can be extrapolated to match similar power spectral bends in stellar mass black hole X-ray binaries \citep{2006Natur.444..730M}. The comparison with black hole X-ray binaries (BHXRBs) also makes the low-frequency power spectral shapes of interest. For example, most AGN show single power spectral bends from high-frequency slopes $\alpha>2$ to low-frequency slopes $\alpha\simeq 1$, similar to those seen in BHXRB soft states \citep{2004MNRAS.348..783M,UttleyMcHardy2005}. A few AGN show evidence for a second, lower-frequency bend, marking a similarity with BHXRB hard and hard-intermediate states \citep{2007MNRAS.378..649S,2007MNRAS.382..985M}. More generally, the broadband power spectral shapes of accreting black holes can be used to constrain the physical origin of the variability process, e.g. in terms of accretion rate variations produced at different radii in the accretion disk \citep{2006MNRAS.367..801A,UttleyMalzac2024}.

The requirement of regular time sampling for the conventional periodogram means its application to AGN data is limited, since such data are usually irregularly sampled. Furthermore, the periodogram is subject to Fourier leakage effects - undesirable distortions and correlated errors between sampled frequencies. For example, steep power spectral slopes at low frequencies lead to so-called `red noise leak' which contributes a bias with slope $\alpha=2$ \citep{2002MNRAS.332..231U} and makes it difficult to accurately constrain steeper power spectral slopes.

Fourier methods using forward-modelling approaches have been proposed to overcome these problems \citep{1992ApJ...400..138D,2002MNRAS.332..231U}. In these approaches, the data are interpolated onto an even grid and the periodogram is computed. This is then compared to the results of simulations based on a model power spectrum: long simulations of time series are generated, resampled to match the observed data, and the interpolation and periodogram computations applied to the simulation. This process is repeated over many simulations to produce an average periodogram to be compared with the data. These methods allow for analysis of irregularly sampled data but, being simulation-based, the estimation of the best-fitting power spectral shape is slow and limited to only a few free parameters. The likelihood function for this is not known, and so a fit statistic  is used for its similarity to standard chi-square fitting. The distribution of the fit statistic is not known; quantifying the goodness of fit and parameter uncertainties requires a large number of simulations, and depends on the choice of fit statistic \citep[see][for a discussion of some issues]{2007MNRAS.378..649S,MuellerMadejski2009,2010ApJ...724...26M}.

In optical AGN variability studies, sampling limitations can be even more significant, e.g. due to weather, telescope scheduling, and seasonal visibility of targets from ground-based observatories. However, unlike X-ray astronomy, optical studies of AGN variability often work in the time domain \citep[e.g. ][]{2013ApJ...765..106Z}. Most time domain methods have the advantage of being free of the biases of Fourier forward-modelling of the power spectrum, but at the expense of assuming a statistical distribution (and hence power spectral form) for the process \citep{1981ApJS...45....1S}, which may make the method unsuitable for modelling observed AGN light curves (e.g. see \citealt{Kozlowski2016}). However, these methods are not biased by irregular sampling and can account for heteroscedasticity when the error bars are different between flux measurements. Moreover, if the likelihood can be calculated in terms of the fitted model parameters, rapid gradient-descent-based fitting approaches are possible which enable rapid modelling of large samples of light curves, for example from massive time-domain surveys.

Over the years, Gaussian processes (GPs) have been widely adopted in time domain astronomy \citep{2023ARA&A..61..329A} and applied to study time series. Using standard Gaussian process regression, \citet{2010MNRAS.403..196M} and \citet{2013ApJ...777...24Z} provided first attempts at modelling flexible power spectra but were computationally expensive. The Damped Random Walk (DRW), a particularly simple GP model, has recently become popular as a model for quasar variability \citep[see e.g.][]{2009ApJ...698..895K}. The power spectrum of a DRW process is a Lorentzian centred at zero frequency, with a high-frequency slope $\alpha=2$, inconsistent with many studies of AGN power spectra in both X-ray \citep{GonzalezMartinandVaughan2012} and optical bands \citep{Edelsonetal2014}.

\cite{2014ApJ...788...33K} introduced Continuous Auto-Regressive Moving Average (CARMA) processes to astronomy. These enable a fast and flexible way of modelling stochastic time series. This power spectrum model can be seen as a weighted sum of modified Lorentzians \citep{vysakhsthesis}. Although this is more flexible than a single Lorentzian, the power spectrum decays with a fixed integer slope depending on the order of the process and flattens at low frequencies. It can also model narrow features such as QPOs on top of the broad-band power spectrum. The parametrisation of this model makes it difficult to properly define the relatively featureless power spectra expected in most models of AGN variability. Proper characterisation of the continuum noise is crucial for defining a null hypothesis when quantifying the significance of periodicities or QPOs in red noise \citep[see][]{2010MNRAS.402..307V,vysakhsthesis}. A limitation of CARMA models for this purpose is that they can implicitly include QPO-like signals in the underlying power spectrum.

A sophisticated Gaussian process time domain model named \texttt{celerite} was presented in \cite{2017AJ....154..220F}. This model allows for steep power spectra in the form of $f^{-4}$ with a fixed integer slope at high frequencies and fast inference. This later tool is widely used for exoplanet science \citep[e.g.][]{2019MNRAS.490.2262E}, and has been used occasionally for variability studies of active galaxies \citep[e.g. ][]{2021ApJ...919...58Z,2023ApJ...946...52Z}. For all of these models, bend frequencies may be flexible but the high-frequency slope is always a fixed integer -- either $-2$ or $-4$ and the low-frequency power spectrum is constrained to be flat.

In this paper we present a new Gaussian process model which aims to avoid all these limitations -- a method that will work on data with arbitrary time sampling, made up of data points which may have very different measurement uncertainties and perhaps even come from different instruments, but allows the use of models for the power spectrum shape that are consistent with our best current understanding of AGN variability. This uses Gaussian process models for their flexibility and ability to model irregular data, providing a well-defined likelihood function that can be used in Bayesian modelling. Our innovation here is to combine the GP framework with a way to approximate the desired power spectral shapes in a way that allows for fast computation. Although developed specifically to aid the study of AGN variability, it can be applied to time series data in general. We make use of flexible bending power-law models for the power spectrum by approximating the model with simple basis functions chosen for their amenability to fast computational methods. This allows proper estimation of the bend frequencies and slopes of the power spectrum with a simpler parametrisation compared to current methods. Building the method with the basis functions of \cite{2017AJ....154..220F} allows fast likelihood computation on large datasets.

In \cref{sec:thepsd} we define formally the power spectrum and the statistical assumptions of this work in the context of astronomical time series and present Gaussian process regression in the general context and current models used in astronomy. \cref{sec:pioran} presents our approach to the power spectrum approximation and how we implement an efficient likelihood evaluation. In \cref{sec:bayesianworkflow}, we present the Bayesian workflow adopted in this work, from the modelling to diagnostics post-inference. We test our method with simulation-based calibration in \cref{sec:simulationbased}. We apply our method to the long-term light curve of the Seyfert 1 galaxy Ark 564 in \cref{sec:ark564} to estimate its power spectrum. We discuss limitations and possible improvements of this method in \cref{sec:discussion}. \cref{sec:conclusion} concludes on the method and results presented in this work.

In order to describe the data, models, associated statistics and their properties, we require a system of notation. Here, we adopt notation for the GP theory drawn from \citet{2006gpml.book.....R}. Specifically, a subscript asterisk is used to denote a test set, e.g. $\boldsymbol{t}$ might be a vector of times where a process has been observed, and $\boldsymbol{t_*}$ is a set of times at which we wish to predict the value of the process. Vector quantities are in bold (e.g. $\boldsymbol{x}$; a column vector), matrices are uppercase Roman or Greek letters (e.g. $K$, $\Sigma$). $K^T$ and $K^{-1}$ are the transpose and inverse of the matrix $K$, respectively, and $|K|$ is its determinant. We use $\sim$ to mean `distributed according to' to indicate the probability distribution of a variable or process. \cref{tab:Notations} summarises the notation and symbols used in this paper.

\begin{table}
    \centering
    \begin{tabular}{lll}
        \toprule\toprule
        \multicolumn{3}{c}{Observation}                                                                                                 \\\midrule
        $\boldsymbol{x}$                                                         & Observed time series                               & \\
        $\boldsymbol{t}$ or $t$                                                  & Time                                               & \\
        $\boldsymbol{\sigma^2}$                                                  & Variance of the measurement process                & \\
        $\Delta t$                                                               & Sampling period or time spacing in the time series & \\
        $T$                                                                      & Duration of the time series                        & \\
        $f_\mathrm{min}=1/T$                                                     & Minimum frequency in the time series               & \\
        $f_\mathrm{max}=1/2\mathrm{min}(\Delta t)$                               & Maximum frequency in the time series               & \\ \\
        \multicolumn{3}{c}{Gaussian process}                                                                                            \\\midrule
        $\boldsymbol{f}$                                                         & Gaussian process                                   & \\
        $\boldsymbol{f_*}| \boldsymbol{t},\boldsymbol{x},\boldsymbol{t_*}$       & Conditioned Gaussian process                       & \\
        $\boldsymbol{t_*}$                                                       & Times for the prediction                           & \\
        $\boldsymbol{x_*}(\boldsymbol{t_*})$                                     & Predicted time series                              & \\ \\

        \multicolumn{3}{c}{Modelling}                                                                                                   \\\midrule
        $\mathcal{R}(\tau)$                                                      & Autocovariance function                            & \\
        $\mathcal{P}(f)$                                                         & Power spectral density                             & \\
        $f$                                                                      & Frequency                                          & \\
        $\tau$                                                                   & Time delay                                         & \\
        $S_{\rm low}$                                                            & Scale factor to extend the low-frequencies         & \\
        $S_{\rm high}$                                                           & Scale factor to extend the high-frequencies        & \\

        $f_\mathrm{start} = f_\mathrm{min} / S_{\rm low}$                        & Minimum frequency                                  & \\
        $f_\mathrm{stop} = f_\mathrm{max}  S_{\rm high}$                         & Maximum frequency                                  & \\

        $J$                                                                      & Number of basis function for the approximation     & \\
        $\psi(f)$                                                                & Basis function in the Fourier domain               & \\
        $\phi(\tau)$                                                             & Basis function  in the time domain                 & \\

        \\

        \multicolumn{3}{c}{Parameters}                                                                                                  \\\midrule
        $\boldsymbol{\theta}$                                                    & Parameters                                         & \\
        $\alpha_i$                                                               & Index/slope of the power-law                       & \\
        $f_{b,i}$                                                                & Bend frequency                                     & \\
        variance                                                                 & $\mathcal{R}(0)$ or integral of $\mathcal{P}(f)$   & \\
        $\nu$                                                                    & Scale factor on the measurement uncertainties      & \\
        $\mu$                                                                    & Mean of the Gaussian time series                   & \\
        $c$                                                                      & Constant to shift a log-normal time series         & \\
        $\gamma$                                                                 & Inter-calibration factor between two time series     \\
        \\
        \multicolumn{3}{c}{Inference}                                                                                                   \\\midrule
        $\boldsymbol{d}$                                                         & Data                                               & \\
        $p(\boldsymbol{d}|\boldsymbol{\theta})=\mathcal{L}(\boldsymbol{\theta})$ & Likelihood function                                & \\
        $p(\boldsymbol{\theta})$                                                 & Prior probability density                          & \\
        $Z$                                                                      & Bayesian evidence                                  & \\
        $BF$                                                                     & Bayes factor                                       & \\
        \bottomrule\bottomrule
    \end{tabular}
    \caption{Symbols used throughout this work.}
    \label{tab:Notations}
\end{table}

\section{Random processes}
\label{sec:thepsd}

If a random process $\{X(t)\}$ is weakly stationary, also known as {\it wide sense stationary} (WSS), it can be fully described by its mean $\mu_X$ and autocovariance function $\mathcal{R}(\tau)$ given by \cref{eq:autocovariance}. The expectation operator $\mathbb{E}$, represents the ensemble mean, the average over all possible realisations of the process $\{X(t)\}$, this is not the time average. From now on, we will assume to be working with random processes that are stationary in the wide sense.
\begin{align}
    \mathcal{R}(\tau) & = {\rm Cov}(X(t),X(t+\tau)) \nonumber                                                                         \\
                      & =\mathbb{E}\left[ \left(X(t)- \mu_X \right) \left(X(t+\tau)- \mu_X \right) \right]. \label{eq:autocovariance}
\end{align}
The autocovariance function of the process quantifies how the values of the time series are correlated with each other when separated in time by $\tau$. It is worth noting that the variance of the process is given by $\mathcal{R}(0)=\mathrm{Var}(X)$ and should always be positive. The Wiener-Khinchin theorem states that the autocovariance function and the power spectrum for frequency $f$, $\mathcal{P}(f)$, are Fourier pairs \citep{priestley,chatfield2004timeseries}:

\begin{equation}
    \mathcal{P}(f) = \int\displaylimits_{-\infty}^{+\infty} \mathcal{R}(\tau) e^{-2{\rm i}\pi f \tau} {\rm d }\tau \;,  \quad \mathcal{R}(\tau) = \int\displaylimits_{-\infty}^{+\infty}  \mathcal{P}(f) e^{2{\rm i}\pi f \tau} {\rm d }f.
    \label{eq:psdacvf}
\end{equation}

For this work, we will consider power spectral densities which yield finite variance and thus must be absolutely integrable. By definition, if the power spectrum is well-behaved, the autocovariance function of the process will be positive semi-definite.

\subsection{Gaussian Processes}
\label{sec:gp}

Here we introduce Gaussian processes which enable the modelling of the variability of Gaussian time series with arbitrary sampling using autocovariance function. Over the last decade, Gaussian processes have been widely used in time domain astronomy with many robust and well-tested codes  -- see the review by \citet{2023ARA&A..61..329A}.

Gaussian processes are a class of random processes for which the joint probability distribution of any finite set of random variables is a Gaussian distribution \citep{2006gpml.book.....R}.  The probability density of a multivariate ($D$-dimensional) Gaussian with mean $\boldsymbol{\mu}$ and covariance matrix $\Sigma$ is given by \cref{eq:densityGP}.
\begin{equation}
    p(\boldsymbol{x}|\boldsymbol{\mu},\Sigma) = \left(2\pi\right)^{-D/2}|\Sigma|^{-1/2}\exp{\left(-\dfrac{1}{2}\left(\boldsymbol{x}-\boldsymbol{\mu}\right)^\intercal \Sigma^{-1} \left(\boldsymbol{x}-\boldsymbol{\mu}\right)\right)}.
    \label{eq:densityGP}
\end{equation}

A Gaussian process $\boldsymbol{f}$, with one-dimensional output, is described by a mean function $\mu(t)$ and a covariance function $k(t,s)$. We will assume the mean function to be constant $\mu(t)=\mu$. As we intend to infer the statistical properties of a time series modelled with a time-symmetric stationary process, the covariance function will depend only on the time separation $\tau = |t-s|$. We write equivalently $\mathcal{R}(\tau)=k(t,t+\tau)$.

\subsubsection{Inference with Gaussian Processes}

\label{sec:likelihood}

When the covariance function is a function of a vector of parameters $\boldsymbol{\theta}$, then we can find the set of `best-fitting' parameters to perform regression. This can be done by either maximising the log-likelihood or in a Bayesian framework sampling a posterior distribution. The log-likelihood function associated with the Gaussian process is given in \cref{eq:likelihood}.

\begin{multline}
    \ln\mathcal{L}(\boldsymbol{\theta},\nu,\mu)=-\frac{1}{2}\left( \boldsymbol{x} - \mu \right)^{\rm T} \left(K +\nu\boldsymbol{\sigma^2} I\right)^{-1} \left( \boldsymbol{x} - \mu \right) \\
    -\dfrac{1}{2}\ln\left|K +\nu\boldsymbol{\sigma^2} I\right| - \dfrac{N}{2}\ln(2\pi).\label{eq:likelihood}
\end{multline}

where $\boldsymbol{x}$ is the vector containing the noisy observed time series with measurement variance $\boldsymbol{\sigma^2}$ at times $\boldsymbol{t}$ and $I$ is the identity matrix. $K=\boldsymbol{k}(\boldsymbol{t},\boldsymbol{t})$ is the covariance matrix obtained by evaluating the covariance function at the observed times. $\mu$ is the mean of the time series, $\nu$ is a scaling parameter on the measurement variance $\boldsymbol{\sigma^2}$. If $\boldsymbol{\sigma^2}$ is a good approximation to the variance of the measurement process $\nu$ should be around one, i.e. the error bars are reliable. If $\nu>1$ then the uncertainties are underestimated, alternatively if $\nu<1$ the uncertainties are overestimated \citep[see e.g.][]{2016MNRAS.461.3145V}. The first term in the log-likelihood can be seen to improve the quality of the fit, the second term penalises complex models and the last term is a normalising constant.

\subsubsection{Prediction}

Given a set of parameters $\boldsymbol{\theta}$, we can also predict the time series at observed and arbitrary times. A Gaussian process can be used to perform regression on a time series $\boldsymbol{x}$, regardless of the sampling, to proceed, one must first choose the covariance function. Gaussian distributions have several convenient properties, the marginalised and conditioned distributions of Gaussian random variables are also Gaussian. Therefore, Gaussian process regression can be seen from a conditional probability point of view as the posterior distribution obtained after conditioning a prior distribution with the data. The prior Gaussian process is only defined by the choice of mean $\mu(t)$ and covariance function. The posterior or conditioned process $\boldsymbol{f_*}$ is Gaussian with mean and covariance matrix given by \cref{eq:conditionmean,eq:conditioncov}.

\begin{align}
    \mathbb{E}[\boldsymbol{f_*}| \boldsymbol{t},\boldsymbol{x},\boldsymbol{t_*}]   & = K_* \left[K +\boldsymbol{\sigma^2} I \right]^{-1} \boldsymbol{x}\label{eq:conditionmean} \\
    \mathrm{Cov}[\boldsymbol{f_*}| \boldsymbol{t},\boldsymbol{x},\boldsymbol{t_*}] & = K_{**} - K_* \left[K +\boldsymbol{\sigma^2} I \right]^{-1} {K_*}^{\rm T}.
    \label{eq:conditioncov}
\end{align}

Where $\boldsymbol{x}$ is the vector containing the noisy observed time series with measurement variance $\boldsymbol{\sigma^2}$ at times $\boldsymbol{t}$ and $I$ is the identity matrix. $K=\boldsymbol{k}(\boldsymbol{t},\boldsymbol{t})$ is the covariance matrix obtained by evaluating the covariance function at the observed times, $K_{**}=\boldsymbol{k}(\boldsymbol{t}_*,\boldsymbol{t}_*)$ is the equivalent for the prediction times $\boldsymbol{t}_*$. $K_*=\boldsymbol{k}(\boldsymbol{t}_*,\boldsymbol{t})$ is the covariance matrix between observed and prediction times. Adding the term $\boldsymbol{\sigma^2}$ to the diagonal of the covariance matrix assumes that all values of $\boldsymbol{\sigma^2}$ are independent and Gaussian-distributed. The vector of measurement errors $\boldsymbol{\sigma^2}$ can account for the individual measurements having different sizes of error bar (i.e. heteroscedastic data).

It is worth noting that the computation of the posterior distribution in \cref{eq:conditionmean,eq:conditioncov} and likelihood in \cref{eq:likelihood} contains a matrix inversion. In practice, a Cholesky decomposition is used to compute this term but the computational cost of this operation scales as $\mathcal{O}(N^3)$ where $N$ is the number of points in the time series. Additionally, the cost for storing the covariance matrix is $N\times N$ which makes regression difficult on large datasets -- when $N>1000$.

\subsection{Covariance functions}

The choice of covariance function (also sometimes called "kernel" in the literature) will directly impact the shape and smoothness of the realisations of the Gaussian process \citep{2006gpml.book.....R}. To be properly defined, the covariance function must be positive definite, which means it yields a positive definite covariance matrix. By definition, if the power spectral density is positive-valued, the covariance function is positive definite. As stated earlier, Gaussian process regression is generally limited to a small number of points but several covariance functions are structured in a way that makes the computations tractable on large datasets. Here we present the main covariance functions used for AGN time-series modelling and their associated power spectral forms.


\subsubsection{Exponential}
The exponential covariance function is associated with the Lorentzian power spectrum ($\psi_2$ in \cref{tab:covgp}). A Gaussian process with this covariance is often referred to as a damped random walk (DRW) in astronomy or an Ornstein-Uhlenbeck process. The high-frequency power spectrum decays as $1/f^2$ while the low-frequency power spectrum is flat.

As said earlier, Gaussian process regression is generally limited to a small number of points but several covariance functions are structured in a way that makes the computations tractable on large datasets.

\subsubsection{CARMA}
\label{sec:carma}
Based on the works of \cite{JONES1981651,jonesackerson,belcher1994}, \cite{2014ApJ...788...33K} introduced continuous autoregressive moving average (CARMA) processes to astronomy for inference of the power spectrum. A CARMA$(p,q)$ process is composed of an autoregressive process of order $p$ and a moving average process of order $q$, where $0\leq q \leq p-1$. This process is defined according to a Stochastic Differential Equation (SDE) and in the assumption of Gaussian noise, it is a Gaussian process with analytical covariance function and power spectrum. The order $(p,q)$ of the SDE relates to the number of parameters and can be arbitrarily large, providing a very flexible modelling. The covariance function is a sum of complex exponentials and the power spectrum can be expressed as a weighted sum of $p$ modified Lorentzians \citep{vysakhsthesis}.

The cost of this method scales linearly with the number of data points, this comes from the state-space representation of the process which enables the use of Kalman recursions \citep{brockwell2016introduction}. By choosing the order $(p,q)$  it is possible to infer the shape of the power spectrum with a flexible sum of Lorentzians. One of the caveats of this method for estimating power spectra is that the number of parameters increases with the order so more parameters will need to be constrained which would make the inference slower.

\subsubsection{Celerite}

\cite{2017AJ....154..220F} generalised the CARMA model to a broader class of covariance functions called \texttt{celerite}. This covariance function is a mixture of $J$ exponentials and trigonometric functions with $4J$ coefficients. With specific constraints on the coefficients, the power spectrum decreases as $\propto f^{-4}$ at high frequencies and is flat at low frequencies. The \texttt{celerite} covariance matrix has a semi-separable structure \citep{https://doi.org/10.1002/nla.2003} which is exploited in a fast and stable algorithm presented in \cite{2017AJ....154..220F}. The GP log-likelihood and the posterior mean of the Gaussian process can be computed in a $\mathcal{O}(NJ^2)$ time. Similarly, the memory cost of the operations is drastically reduced as the full covariance matrix is not saved in the memory. Finally, thanks to the convenient properties of semi-separable matrices, the product and the sum of \texttt{celerite} covariance functions are also semi-separable. A specific covariance function of interest for this work is the SHO covariance function associated with a stochastically driven damped simple harmonic oscillator. Its application is presented in \cref{sec:approx}.

\section{\texttt{pioran}: a power spectrum inference method}
\label{sec:pioran}
We are interested in inferring the statistical properties of the underlying process generating the variability observed in AGN light curves. To do so, we want to estimate the power spectrum of AGN. As Fourier methods can be biased and limited to regular data we will work in the time domain using Gaussian processes. We will make use of the covariance functions presented in the previous section to build a general and scalable method to infer the shape of the power spectrum of AGN light curves.

We present \texttt{pioran}\footnote{Power spectrum Inference Of RANdom time series} a method to infer the broad-band power spectrum of random time series using Gaussian processes.

\subsection{Model for the continuum variability}
\label{sec:bendingpowerlaw}
The continuum power spectrum of accreting black holes is well modelled with power-laws and Lorentzians \citep[e.g.][]{2007MNRAS.382..985M}. These phenomenological models are based on the examination of many periodograms of AGN and X-ray binary time series from previous work. In this work, we model the power spectrum with a bending power-law model \citep{2004MNRAS.348..783M,2007MNRAS.378..649S}, with $n$ bends located at frequencies  $f_i$  and $n+1$ slopes $\alpha_i$.

\begin{equation}
    \mathcal{P}(f) = A\left(\frac{f}{f_{b,1}}\right)^{-\alpha_1}\prod_{i=1}^{n} \left[{1+(f/f_{b,i})^{(\alpha_{i+1}-\alpha_{i})}}\right]^{-1}.
    \label{eq:modelPL}
\end{equation}
Here, we only use up to $n=1-2$ bends. When $\alpha_1 \simeq 0$ and $\alpha_2 \simeq 2$, this model is similar to a Lorentzian centred at zero and the process reduces to a damped random walk (DRW). To have a finite integrated power, the low-frequency slope must be less than one, i.e. $\alpha_1<1$ and the high-frequency slope $\alpha_{n+1}$ must be steeper than one, i.e. $\alpha_{n+1}>1$. Another way to ensure finite power is to use cut-offs to limit power at low and high frequencies.
To use this flexible model in a Gaussian process framework, one has to compute the associated autocovariance function with the inverse Fourier transform. Unfortunately, there is no known analytical Fourier transform for such a model. We therefore cannot write a close form expression for the autocovariance function needed for \cref{eq:likelihood}.
We can only rely on approximate methods. Here we present a method for approximating the autocovariance function and using it in the Gaussian process regression.

\subsection{The approximation}
\label{sec:approx}

We approximate our power spectrum model, $\mathcal{P}(f)$ in \cref{eq:modelPL}, using a finite set of simple functions as shown in \cref{eq:approx}.
\begin{align}
    \tilde{\mathcal{P}}(f) & =  \sum\limits_{j=0}^{J-1} a_j \psi(f/f_j)\label{eq:approx}
\end{align}

We call the functions $\psi(f)$ "basis functions", although these do not form a proper basis. They can approximate a certain range of power spectral shapes, but not all valid power spectra can be expressed as a sum of these functions. The inverse Fourier transform of these basis functions forms an approximation to the autocovariance model $\mathcal{R}(\tau)$.

\begin{align}
    \tilde{\mathcal{R}}(\tau) & = \sum\limits_{j=0}^{J-1} a_j f_j \phi(\tau f_j)
    \label{eq:acvf_approx}
\end{align}

We use four requirements to inform the choice of basis functions, $\psi(f)$, they should:
\begin{enumerate}
    \item be smooth in the frequency domain;
    \item have a power-law shape at lowest and highest frequencies;
    \item have a known autocovariance function;
    \item have an autocovariance function amenable to fast computation methods.
\end{enumerate}

Requirements (i) and (ii) ensure that the sum of a relatively lower number of basis functions can approximate well a smooth, broad-band spectrum with power-law form at the highest and lowest frequencies. Requirement (iii) enables us to apply GP regression (\cref{eq:likelihood}), and requirement (iv) allows us to apply these methods efficiently. The \texttt{celerite} \citep{2017AJ....154..220F} covariance functions will satisfy these requirements. In \cref{tab:covgp}, we present the basis functions $\psi$ used for the approximation and their associated inverse Fourier transform $\phi(\tau)$. $\psi_4$ is the power spectrum associated with a stochastically driven damped simple harmonic oscillator (SHO) when the quality factor equals $Q=1/\sqrt{2}$ \citep{2017AJ....154..220F}.  $\psi_6$ can be obtained using partial fraction decomposition and recognising it to be the sum of a Lorentzian and a general \texttt{celerite} power spectrum, this is detailed in \cref{apdx:fourier}.

\begin{table*}
    \caption{Basis functions and their autocovariance function for Gaussian process regression. \textbf{References}: (1) \citet{2006gpml.book.....R}, (2) \citet{2017AJ....154..220F} and (3) this work.}
    \begin{tabular}{llll}
        \toprule
        Name         & Basis function $\psi(f)$                     & Associated autocovariance function $\phi(\tau)$                                                                                                                                             & Ref. \\\midrule
        DRW          & $\psi_2(f)=\dfrac{A}{\alpha^2 + 4\pi^2 f^2}$ & $\phi_2(\tau)= \dfrac{A}{2\alpha} \exp{\left(- \alpha |\tau|\right)} $                                                                                                                      & (1)  \\
        SHO          & $\psi_4(f) = \dfrac{1}{1+f^4}$               & $\phi_4(\tau) = \dfrac{\pi}{\sqrt{2}}\exp\left(-\pi\sqrt{2}|\tau|\right) \left(\cos\left(\pi\sqrt{2}|\tau|\right)+\sin\left(\pi\sqrt{2}|\tau|\right)\right)$                                & (2)  \\
        DRW+Celerite & $\psi_6(f) = \dfrac{1}{1+f^6}$               & $\phi_6(\tau) =\dfrac{\pi}{\sqrt{3}} \left(\dfrac{ \exp{(-2\pi |\tau|)}}{\sqrt{3}}+\exp{(-\pi|\tau|)}\left(\dfrac{\cos(\pi\sqrt{3}|\tau|)}{\sqrt{3}}+\sin(\pi\sqrt{3}|\tau|)\right)\right)$ & (3)  \\
        \\\bottomrule
    \end{tabular}
    \label{tab:covgp}
\end{table*}
Assuming we can approximate our power spectrum model as in \cref{eq:approx}, the approximated autocovariance function is given by \cref{eq:acvf_approx} where  $a_j$ are real coefficients and characteristic frequencies $f_j$ of the basis functions. $J$ is the number of basis functions. It should be stressed that due to the shape of the basis functions, the approximated model $\tilde{\mathcal{P}}(f)$ will be flat at low frequencies - slope of $0$ - and steep at high frequencies - either a slope of $-4$ or $-6$ depending on whether $\psi_4$ or $\psi_6$ was used. This implies that the integrated power (variance) will always be finite, without the need to introduce additional cut-offs.

\begin{figure}
    \centering
    \includegraphics[width=.5\textwidth]{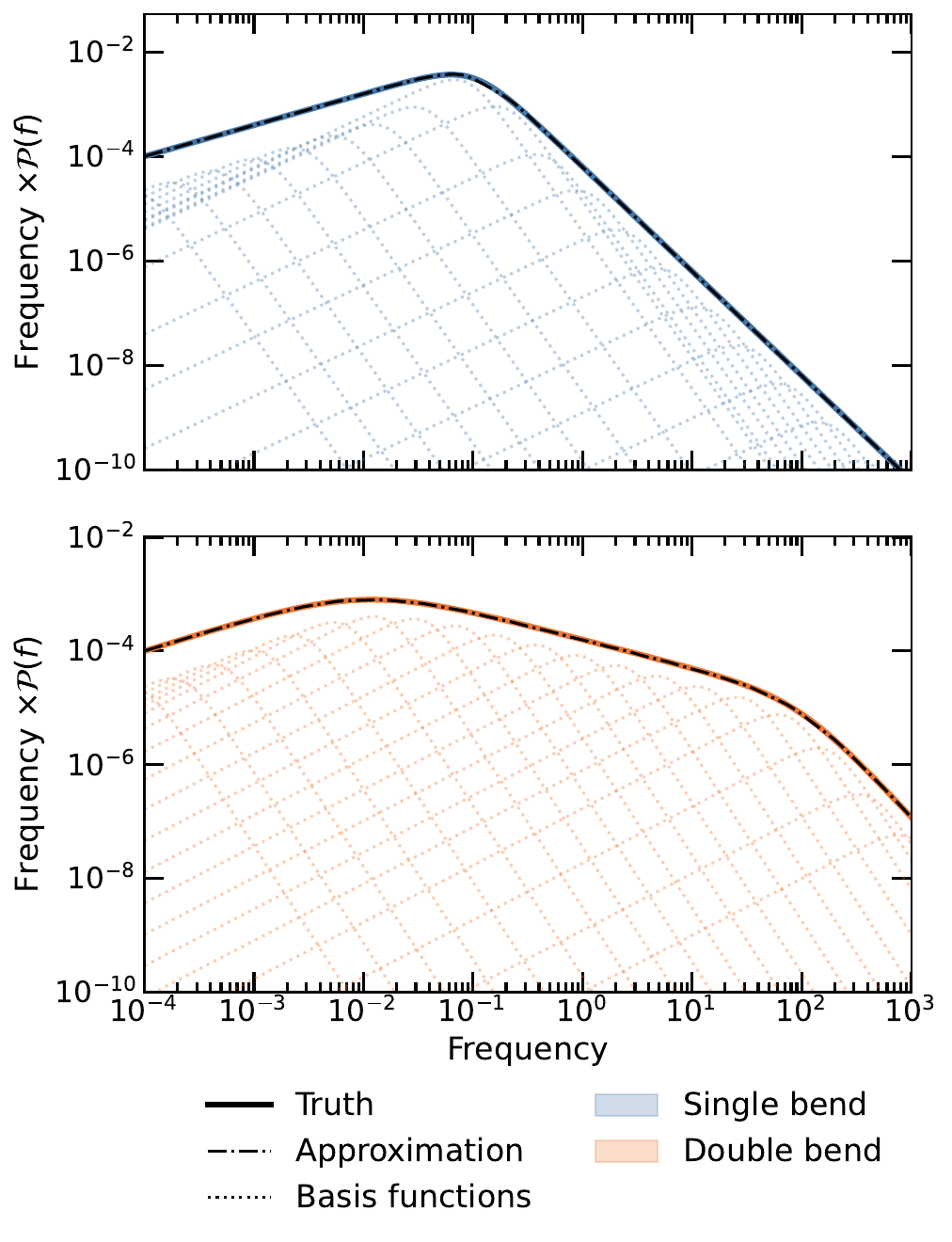}
    \caption{True (solid line) and approximated (dashed-dot line) models with the basis functions (dashed lines) for a single bending (blue) and double bending (orange) power spectrum model.}
    \label{fig:psd_approx}
\end{figure}

The frequencies $f_j$ are geometrically spaced and thus given by: $f_j=f_\mathrm{start}\left({f_\mathrm{stop}}/{f_\mathrm{start}}\right)^{j/(J-1)}$ for $j=0,1,\dots,J-1$. We introduce two scale factors $S_\mathrm{low}$ and $S_\mathrm{high}$ to extend the low and high frequencies from $f_\mathrm{start}=f_\mathrm{min}/S_\mathrm{low}$ to $f_\mathrm{stop}=f_\mathrm{max} S_\mathrm{high}$. In practice, we choose $S_\mathrm{low}=100$ and $S_\mathrm{high}=20$. The constraint $\tilde{\mathcal{P}}(f_j) = \mathcal{P}(f_j)$, forms a system of $J$ linear equations where the unknowns are the $a_j$. The matrix $B$ of dimensions $(J, J)$ representing this system is presented in \cref{eq:approx_system} and one can notice it has a Toeplitz form.

\begin{align}
    \boldsymbol{p} = \boldsymbol{a} B \quad \text{where } B_{ij}=\psi(f_i/f_j) \text{ and } p_j = \mathcal{P}(f_j)
    \label{eq:approx_system}
\end{align}
Due to the small size of the $B$ matrix $J\leqslant50$, standard methods can be used to solve the system. In \cref{apdx:condition}, we find that the matrix can be well-conditioned when the ratio of minimal and maximal frequencies spans several orders of magnitudes. Otherwise, one might need to use fewer basis functions. The approximation in the Fourier domain can be visualised in \cref{fig:psd_approx}.

Once the $a_j$ and $f_j$ are obtained, they can be used in the \texttt{celerite} algorithm \citep{2017AJ....154..220F} to describe the covariance functions and perform regression with a computational time scaling linearly with the number of data points.

\subsubsection{Checking the approximation}
\label{sec:checks}

Before running any inference with this method, one should check that the approximation is accurately describing the model. To check this, we compute residuals and ratios between the true model $\mathcal{P}$ and the approximated model $\tilde{\mathcal{P}}$.

\begin{figure}
    \centering
    \includegraphics[width=\columnwidth]{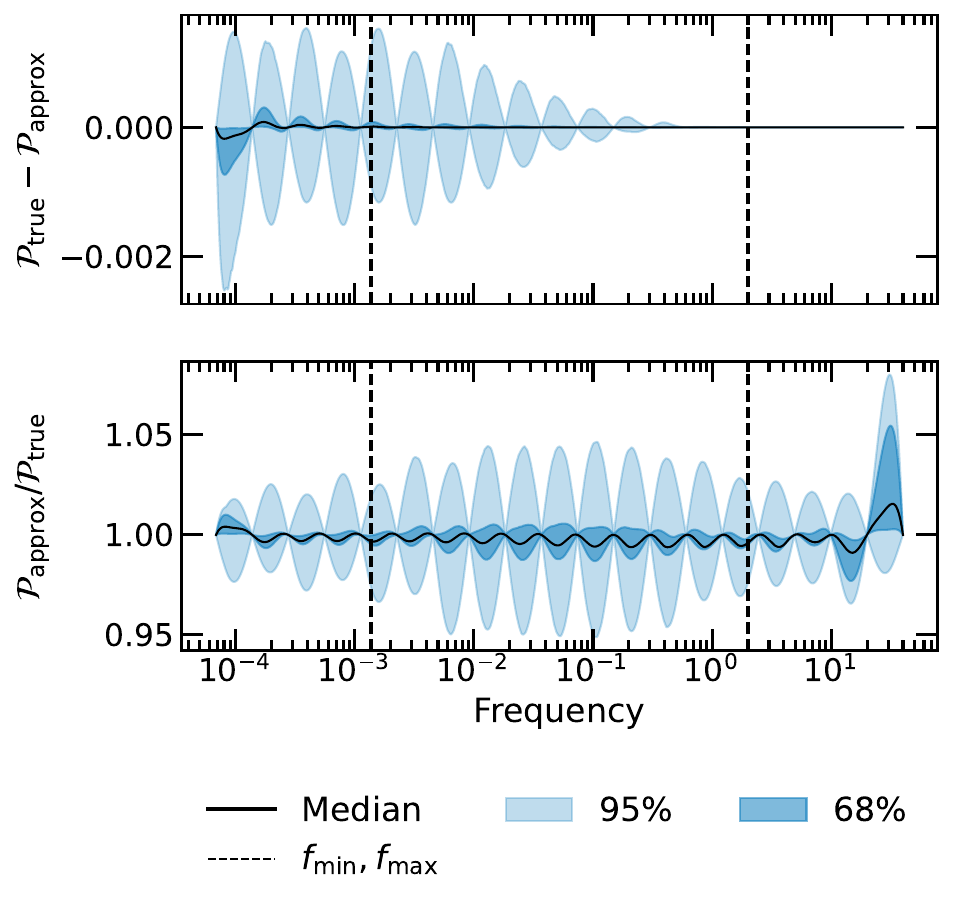}
    \caption{Residuals (top) and ratios (bottom) between the intended power spectrum model and its approximation as a function of frequency. The median, the  $68^\mathrm{th}$ and $95^\mathrm{th}$ percentiles are also shown. The minimum and maximum frequencies of the observed data are shown respectively as black dashed-dot and dotted lines.}
    \label{fig:diagnostics_frequencies}
\end{figure}

First, we draw realistic values for the parameters of the intended model (bending power-law), this can be done by sampling from a prior distribution (see \cref{sec:psdparms} for more details about the priors). Then, for each set of parameters, we compute the power spectrum model and its approximation over a grid of frequencies to obtain a distribution of the intended and the approximated power spectral shapes. The distribution of the residuals and ratios are respectively given by the difference or ratio between $\mathcal{P}$ and $\tilde{\mathcal{P}}$. We assess the quality of the approximation by computing quantiles on the distribution of the residuals and ratios; an example is shown in \cref{fig:diagnostics_frequencies} where we approximate the model with $J=20$ basis functions. In this example, we see that the approximation holds in the range of observed frequencies.

\begin{figure}
    \centering
    \includegraphics[width=\columnwidth]{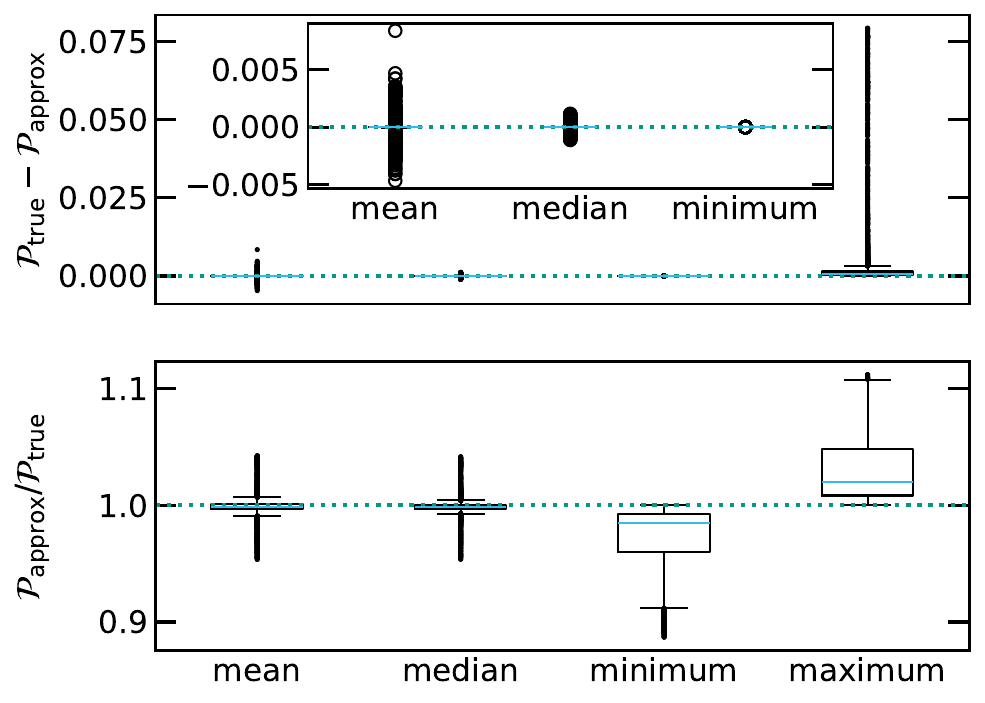}
    \caption{Distributions of the mean, median, minimum and maximum of the frequency-averaged residuals (top) and ratios (bottom) between the true power spectrum model. The extreme values of each distribution are shown as whiskers.}
    \label{fig:diagnostics_metrics}
\end{figure}

We also assess the approximation using other metrics such as the mean, median, maximum, and minimum values of the frequency-averaged residuals and ratios. This is shown in \cref{fig:diagnostics_metrics}, the residuals are mainly located around zero while the ratios are centred around one. A more detailed analysis of the choice of $J$ is given in \cref{apdx:quality}. The quality of the approximation depends on the model approximated, the basis function $\psi$, the number of basis functions $J$ and the size of the frequency grid.

\subsection{Other assumptions}

\label{sec:logtransform}
We have assumed that the time series consists of Gaussian measurements. However, the distribution of flux in accreting black hole light curves appears to be log-normal \citep{2005MNRAS.359..345U}. This is thought to originate from multiplicative independent and identically distributed processes rather than additive processes. This can be observed as a linear relation between the mean flux and the root-mean-square amplitude of variability \citep{2001MNRAS.323L..26U,2005MNRAS.359..345U}. While the origin and interpretation of such a relation is debated \citep{2020ApJ...895...90S}, we only intend to include a way to model the non-Gaussian distribution of fluxes. Assuming the distribution of fluxes is log-normal, taking the logarithm of the observed light curve $x$ makes the data Gaussian as shown in \cref{eq:logtransform}. The additional parameter $c$ accounts for a possible shift in the log-normal distribution. $c$ could be intrinsic to the source variability or instrumental, e.g. background.

\begin{equation}
    y = \ln{\left(x-c\right)} \quad \quad \text{where }\quad 0 \leq c < \min{x}
    \label{eq:logtransform}
\end{equation}

Additionally, the transformation is propagated on the measurement uncertainties as shown in \cref{eq:logtransform_err}. The time series $y$ can now be used in the Gaussian process regression with measurement uncertainties $\sigma_y$.
\begin{equation}
    \sigma_{y}^{2} = \frac{\sigma_{x}^2}{\left(x-c \right)^2}
    \label{eq:logtransform_err}
\end{equation}
Other transformations such as the Box-Cox transformation \citep{boxcox} could be considered to make the data normally distributed. Implicit in the assumption of a Gaussian time series with independent Gaussian measurement errors, we also assume that no outliers are present in the data.\\

If the time series is a combination of times series from multiple instruments, it is possible to include a cross-calibration factor. In \cref{sec:ark564}, we apply the method to real data from two different instruments and use a scale factor named $\gamma$ on one of the time series. $\gamma$ should be one when the cross-calibration is accurate.

\subsection{Implementation}

We implement the method in various packages, the code is available in the Python library \texttt{stingray}\footnote{\url{https://docs.stingray.science/en/latest/}} \citep{Huppenkothen2019,2019ApJ...881...39H} using \texttt{tinygp}\footnote{\url{https://tinygp.readthedocs.io/en/stable/}} for the Gaussian process regression with quasi-separable covariance functions using Just-in-time compilation with JAX \citep{jax2018github}.

\begin{figure}
    \centering
    \includegraphics[width=\columnwidth]{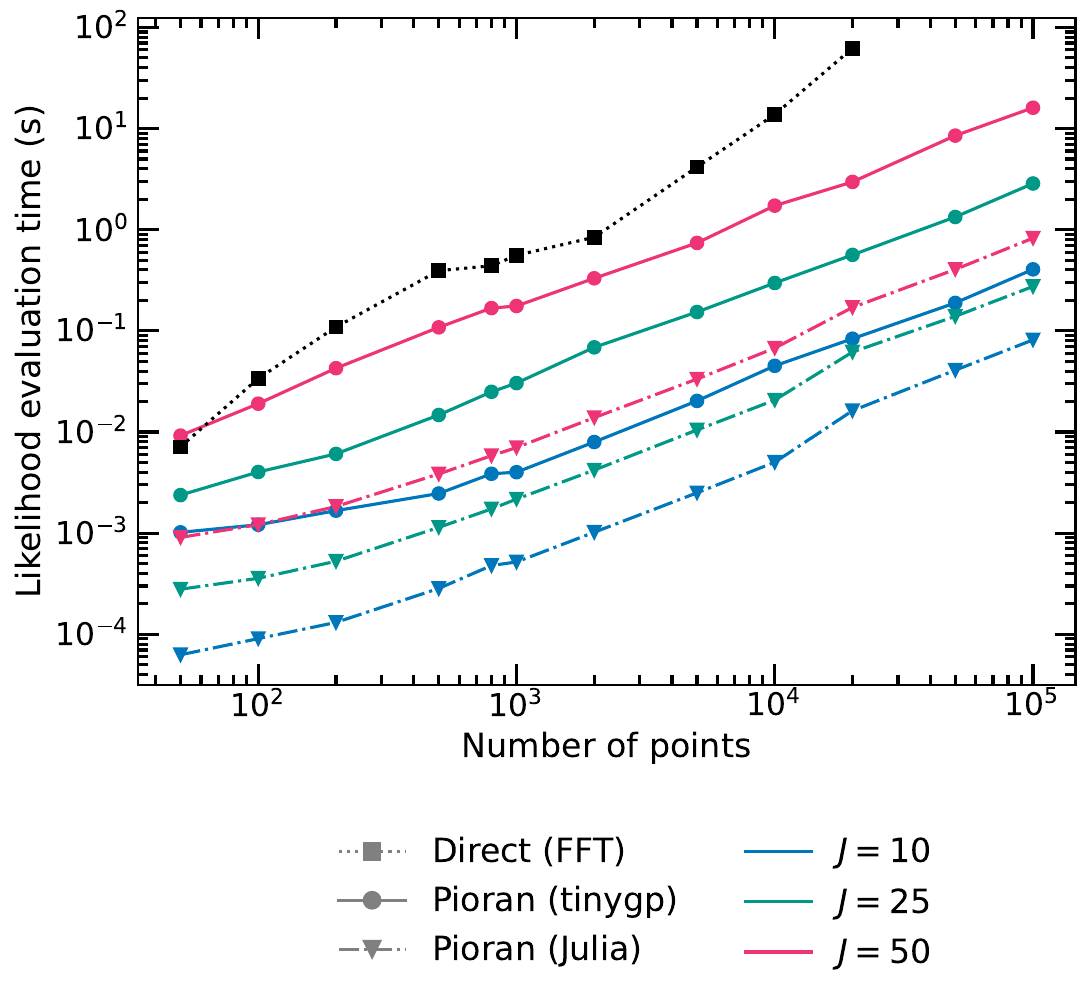}
    \caption{Likelihood evaluation time in seconds for the approximation method and the direct method as a function of the number points and number of basis functions $J$ for the approximation. The black squares show regression with direct method scaling as $\mathcal{O}(N^3)$ (see \cref{sec:fftmethod}). The circles and triangles respectively represent the Python \texttt{tinygp} and  Julia \texttt{Pioran.jl} codes where colours encode the number of SHO basis functions.}
    \label{fig:cost}
\end{figure}

A pure Julia implementation is also available in \texttt{Pioran.jl}\footnote{\url{https://github.com/mlefkir/Pioran.jl}}. In \cref{fig:cost}, we compare the likelihood evaluation time between the implementations in Python and Julia, and also an FFT-based implementation method (see \cref{sec:fftmethod}).

The Python implementation using \texttt{tinygp} appears to be an order of magnitude slower than the Julia implementation but is faster than the direct method. This speed-up is also observed when comparing the C++ implementation of \texttt{celerite}\footnote{\url{https://celerite2.readthedocs.io}} to \texttt{tinygp}.

\section{Bayesian workflow}
\label{sec:bayesianworkflow}
In this work, we adopt a Bayesian workflow \citep{2020arXiv201101808G} for our modelling and to obtain estimates and credible intervals for the parameters of our models. The set of stages proposed in the workflow cover model building, inference, and checks before and after inference, all of this in a Bayesian framework. In \cref{sec:likelihood} we defined the Gaussian process likelihood function, here we define the priors and methods we use in a typical workflow.

\subsection{Bayes rule}
Bayes's rule in \cref{eq:bayesrule} gives the posterior probability density $p(\boldsymbol{\theta}|\boldsymbol{d})$ of parameters $\boldsymbol{\theta}$, knowing that we observed data $\boldsymbol{d}$. The likelihood function is denoted by $p(\boldsymbol{d}|\boldsymbol{\theta})=\mathcal{L}(\boldsymbol{\theta})$ and the prior probability density is given by $p(\boldsymbol{\theta})$.

\begin{equation}
    p(\boldsymbol{\theta}|\boldsymbol{d})  = \dfrac{p(\boldsymbol{d}|\boldsymbol{\theta}) p(\boldsymbol{\theta})}{ p(\boldsymbol{d})}.
    \label{eq:bayesrule}
\end{equation}

The normalising constant in \cref{eq:bayesrule} is called the marginal likelihood, Bayesian evidence or simply evidence and is given by:
\begin{equation}
    \label{eq:marginallikelihood}
    Z = p(\boldsymbol{d}) =\int p(\boldsymbol{d}|\boldsymbol{\theta}) p(\boldsymbol{\theta}) \mathrm{d}\boldsymbol{\theta}.
\end{equation}

The Bayesian evidence can be used to compare two models using the Bayes factor $BF_{12}=Z_1 / Z_2$. When $BF>10$ one model can be favoured, with $BF>100$ the model is strongly supported \citep{1939thpr.book.....J,2017pbi..book.....B}. Sampling from the posterior probability distribution can achieved with Markov Chain Monte Carlo (MCMC) algorithms but to estimate the evidence at the same time, nested sampling algorithms are required \citep{2004AIPC..735..395S}.

\subsection{Priors}
\label{sec:priors}
Before performing any inference on a given time series, we first select a model. Then, we define a joint prior distribution which accounts for any information on the parameters of the model regardless of the data. \cref{tab:modelling} lists all the parameters used for the inference of a power spectrum model with two bends assuming a log-normal distributed time series.

\begin{table*}
    \centering
    \caption{Parameters and priors for the modelling of a power spectrum with two bends. The first half lists the power spectrum parameters and the second half lists the Gaussian process parameters.}
    \begin{tabular}{llllc}
        \toprule\toprule
        Modelling           & Parameter         & Description                                      & Prior distribution                                           & Unit                       \\\midrule
        Power spectrum      & ${\alpha}_1$      & Low-frequency slope                              & Uniform$[0,1.25]$                                            & -                          \\
                            & ${\alpha}_2$      & High-frequency slope or intermediate slope       & Uniform$[\alpha_1,\alpha_{\rm max}]$                         & -                          \\
                            & ${\alpha}_3$      & High-frequency slope                             & Uniform$[\alpha_2,\alpha_{\rm max}]$                         & -                          \\
                            & $f_{b,1}$         & Low-frequency bend                               & Log-uniform$[f_\text{start},f_\text{stop}]$                  & same as $1/t$              \\
                            & $f_{b,2}$         & High-frequency bend                              & Log-uniform$[f_{b,1},f_\text{stop}]$                         & same as $1/t$              \\ \midrule

        Gaussian process or & $\text{variance}$ & Variance of the process                          & Log-normal$(-3,2)$                                           & same as $\boldsymbol{x}^2$ \\
        Time series         & $\nu$             & Scaling on the measurements uncertainties        & Gamma$(2,0.5)$                                               & -                          \\
                            & $\mu$             & Mean of the Gaussian time series                 & Normal$(\bar{x},\beta {s}^2)$                                & same as $\boldsymbol{x}$   \\
                            & $c$               & Offset for a log-normal-distributed time series  & Log-uniform$\left[10^{-6},0.99\min{(\boldsymbol{x})}\right]$ & same as $\boldsymbol{x}$   \\
                            & $\gamma$          & Inter-calibration factor for the two time series & Log-normal(-0.1,0.2)                                         & -                          \\

        \bottomrule        \bottomrule
    \end{tabular}
    \label{tab:modelling}
\end{table*}

\subsubsection{Power spectrum parameters}
\label{sec:psdparms}
In the case of power-law power spectra, our prior beliefs are that the slopes must agree on a decreasing power spectrum. To ensure that the approximation holds, the power spectrum model must not be steeper than $f^{-\alpha_{\rm max}}$ where $\alpha_{\rm max}$ equals $4$ in the case $\psi_4$ and $6$ for $\psi_6$. We also expect the power spectrum to be flat at low frequencies. As shown in \cref{tab:modelling}, we use uniform priors on the slopes, where the slope $\alpha_{i+1}$ depends on the previous slope $\alpha_i$. This is equivalent to defining first $\alpha_1$ with a uniform prior $p(\alpha_1)$, and the prior distribution on $\alpha_2$ is conditional on $\alpha_1$ and then the prior distribution of $\alpha_3$ is conditional on $\alpha_2$ and so on. This can be expressed with $p(\alpha_1)$ and $p(\alpha_2 | \alpha_1)$ in the case of a single-bending power-law.

We require the low-frequency bend $f_1$ to be in a sensible frequency range given by $f_\mathrm{start}=f_\mathrm{min}/S_\mathrm{low}$ and $f_\mathrm{stop} = S_\mathrm{high}f_\mathrm{max}$, where $f_\mathrm{min},f_\mathrm{max}$ denote the limiting frequencies of the time series and $S_\mathrm{low},S_\mathrm{high}$ are scale factors for the approximation. We use a log-uniform prior for the first bend, and similarly to the slopes, the prior on $f_2$ depends on the $f_1$.

We do not model the amplitude of the power spectrum; instead, we include its total integral (from $f=-\infty$ to $f=+\infty$), which is the amplitude of the autocovariance function, as a parameter.

\subsubsection{Gaussian process parameters}

\label{sec:priorsGP}
To specify the prior on the mean of the time series, we extract a random subset of the total time series  -- between one and three per cent of the points --  which we use to compute a sample mean $\bar{x}$ and sample variance $s^2$. This subset is then discarded for the inference and the remainder of the analysis.

If the time series is assumed to be Gaussian then the prior for the mean can be set as a normal distribution centred on $\bar{x}$ with variance $\beta s^2$ where $\beta$ is a scale factor. If we assume the time series to be log-normally distributed, then the values of $\bar{x}$ and $s^2$ are computed using the logarithm of the subset time series.\\

The amplitude of the autocovariance function is the variance of the process and should not be confused with the sample variance of the time series which is an estimator of the variance of the process.
To specify its prior distribution in the context of accreting supermassive black holes, we use $F_\text{var}$ the fractional root mean square variability amplitude defined in \cref{eq:fvar}, where $\overline{\sigma^2_\text{err}}$ is the mean square error \citep{2003MNRAS.345.1271V}.

\begin{equation}
    F_\text{var} = \sqrt{\dfrac{s^2 - \overline{\sigma^2_\text{err}}}{\bar{x}^2}}
    \label{eq:fvar}
\end{equation}

When looking at previous analyses of time series from active galaxies \citep[e.g.][]{2003ApJ...593...96M,2003ApJ...598..935M}, $F_\text{var}$ is almost always less than one. We choose to model $F_\text{var}$ with a log-normal distribution with parameters $\mu_{F_\text{var}}=-1.5$ and $\sigma^2_{F_\text{var}}=1/{2}$. Given that the product of log-normal variables is also log-normal, then $F_\text{var}^2$ also follows a log-normal distribution with mean $2\mu_{F_\text{var}}$ and variance $4\sigma^2_{F_\text{var}}$. To define a prior for the variance of the underlying process we assume that $F_\text{var}^2$ is proportional to the true variance of the process and thus assume that the distribution of $F_\text{var}^2$ can be used as a prior knowledge on the variance. Therefore, we choose a log-normal distribution with parameters $\mu=-3$ and $\sigma^2 = 2$.\\
We choose a $\text{Gamma}(2,0.5)$ prior on  $\nu$ as we expect its value to be close to 1 on average. The Gamma distribution is parametrised with the shape and scale parameters. In the case of a logarithmic transformation of the data (\cref{eq:logtransform}), the offset $c$ in the logarithm has a log-uniform prior. In the case where $c=0$, one can use a half-Cauchy prior as used in \cref{sec:ark564} when applying the method to Ark 564.\\

As mentioned in \cref{sec:approx}, it is important to check that adequate priors were chosen for the approximation to hold, this is a prior predictive check. One can also draw samples from the Gaussian process using these priors to ensure that the time series have realistic values for the mean and variance.

\subsection{Inference}
\label{sec:inference}
In this work, we use the two methods presented below to sample from the posterior probability distributions. As we use \texttt{Pioran.jl}, the Julia implementation of the power spectral approximation, the sampling codes presented here interface well with \texttt{Pioran.jl}.

\subsubsection{Nested sampling}

Nested sampling (NS) enables global parameter exploration and has a natural self-convergence criterion \citep{2004AIPC..735..395S}. Most NS algorithms only require the likelihood function and the joint prior distribution to be evaluated, i.e. the likelihood does not need to be differentiable. In this work, we use \texttt{UltraNest}\footnote{\url{https://johannesbuchner.github.io/UltraNest/}} \citep{2021arXiv210109675B} a robust Python nested sampling code which implements the MLFriends algorithm \citep{2019PASP..131j8005B,2014arXiv1407.5459B}. This implementation requires little tuning and can be parallelised on multiple cores with MPI. We use the default values of 400 live points and $\texttt{frac\_remain} = 0.01$ to obtain samples from the posterior distribution and an estimate of the evidence. Convergence is reached until the target criteria are fulfilled, i.e. the effective sample size is at least $400$ and the new samples contribute to less than $1\%$  to the evidence ($\texttt{frac\_remain} = 0.01$). In practice, to call \texttt{UltraNest} in Julia, we use the package \texttt{PyCall.jl} and \texttt{MPI.jl} \citep{Byrne2021} to speed up the inference with parallelisation.

\subsubsection{Hamiltonian Monte Carlo}

Hamiltonian Monte Carlo (HMC) \citep{2017arXiv170102434B} with the No-U-Turn sampler \citep{2011arXiv1111.4246H} is an MCMC method that provides robust estimates of the posterior samples with little tuning. This method requires the posterior density to be differentiable which is the case in the Gaussian processes implementations of \texttt{tinygp} (Python) and \texttt{Pioran.jl} (Julia). We use the No-U-Turn sampler implemented in the Julia package \texttt{AdvancedHMC.jl} \citep{xu2020advancedhmc} through the probabilistic programming library \texttt{Turing.jl} \citep{ge2018turing}.

By default, we sample $2000$ points from the posterior with a warm-up phase of $2500$ iterations. We sample the posterior with $12$ chains in parallel to use convergence diagnostics such as $\hat{R}$ and the effective sample size (ESS) \citep{Vehtari_2021}. For all parameters, $\hat{R}$ should be close to unity and it is recommended that the ESS should be large enough (e.g. about $\sim 400$). Furthermore, a visual inspection of the chains and the posterior samples can be performed as an additional check for convergence.

\subsection{Diagnostics}

\label{sec:diagnostics}
\begin{figure}
    \centering
    \includegraphics[width=.5\textwidth]{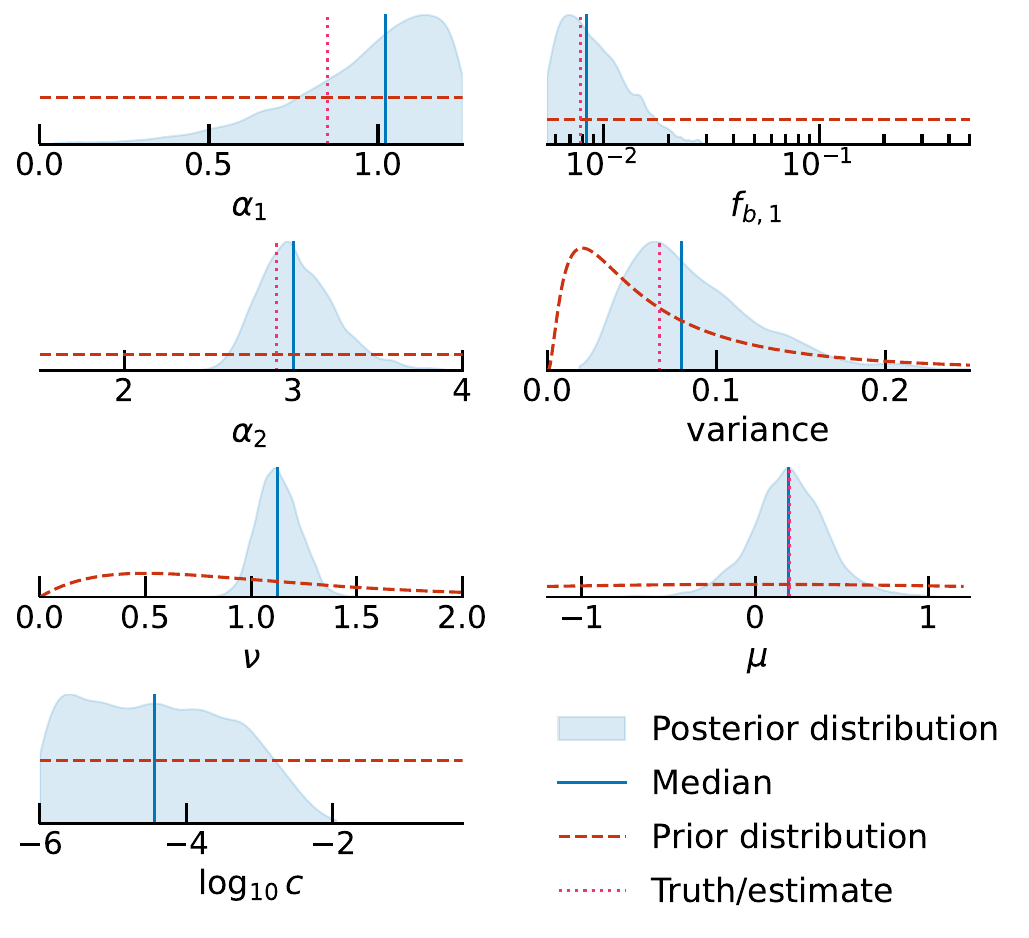}
    \caption{Distributions of the posterior samples for one simulated time series. The median of the distributions is shown by a blue vertical line while the true value is shown by a dotted magenta vertical lines. The prior distributions are shown with dashed red lines.}
    \label{fig:posterior}
\end{figure}

To illustrate the diagnostics in this Section, we use the results of one of the simulations presented in \cref{sec:simulationbased}. Once the posterior samples have been generated, we plot the distribution of the posterior samples against the prior distribution as shown in \cref{fig:posterior}. This allows checking that the priors are not too restrictive, we also compute the median of the distribution as shown by the vertical blue lines in the Figure.

\begin{figure*}
    \centering
    \begin{subfigure}[b]{0.475\textwidth}
        \centering
        \includegraphics[width=\textwidth]{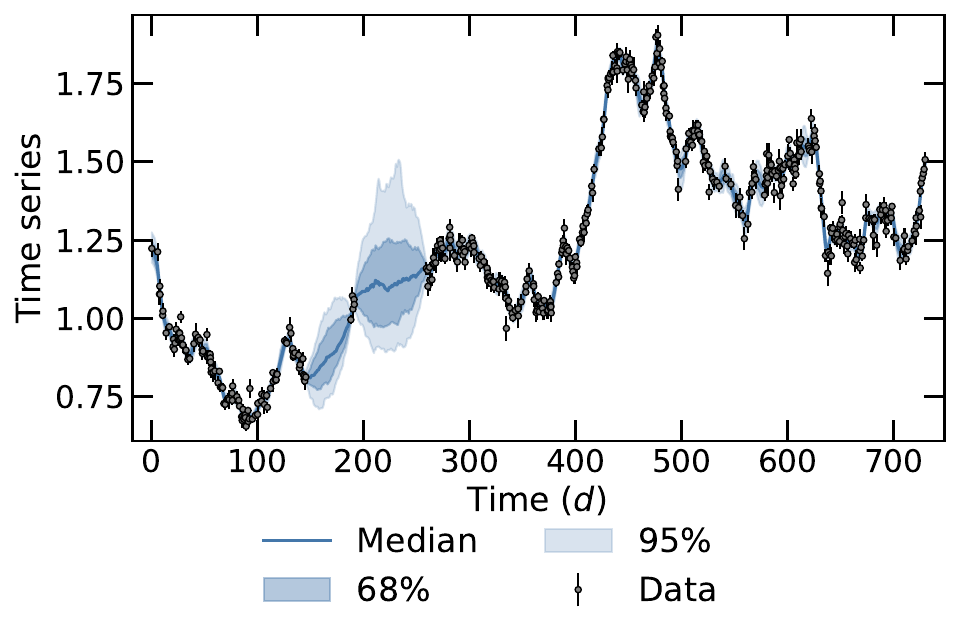}
        \caption{Simulated predictive time series against the observed time series.}
        \label{fig:diag_pred}
    \end{subfigure}
    \hfill
    \begin{subfigure}[b]{0.475\textwidth}
        \centering
        \includegraphics[width=\textwidth]{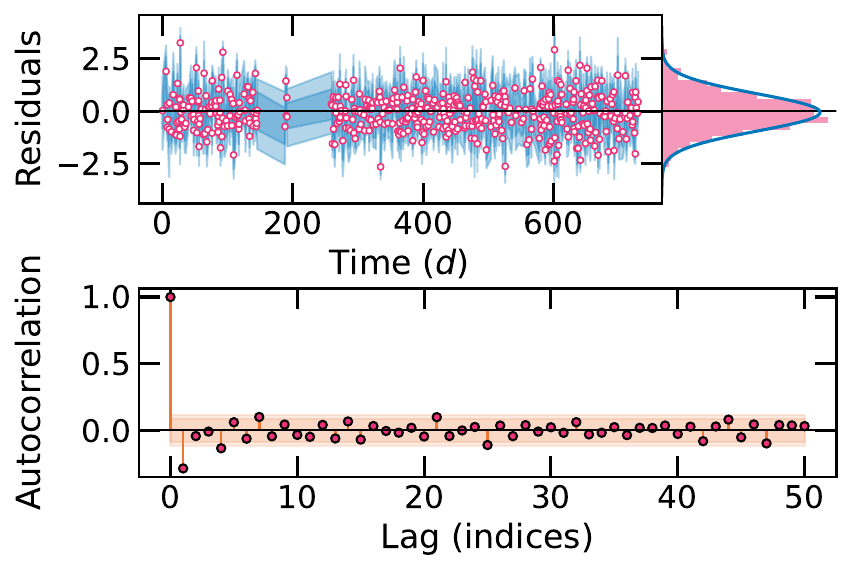}
        \caption{Time series and distribution of the standardised residuals (top panel) and autocorrelation of the residuals (bottom panel).}
        \label{fig:diag_res}
    \end{subfigure}
    \vskip\baselineskip
    \begin{subfigure}[b]{0.475\textwidth}
        \centering
        \includegraphics[width=\textwidth]{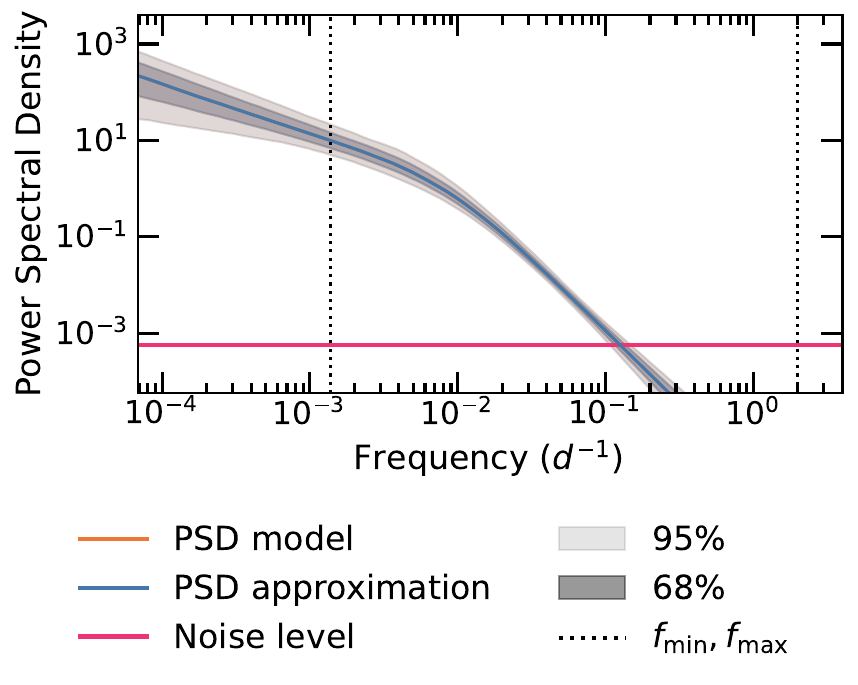}
        \caption{Posterior predictive power spectrum.}
        \label{fig:diag_ppc_psd}
    \end{subfigure}
    \hfill
    \begin{subfigure}[b]{0.475\textwidth}
        \centering
        \includegraphics[width=\textwidth]{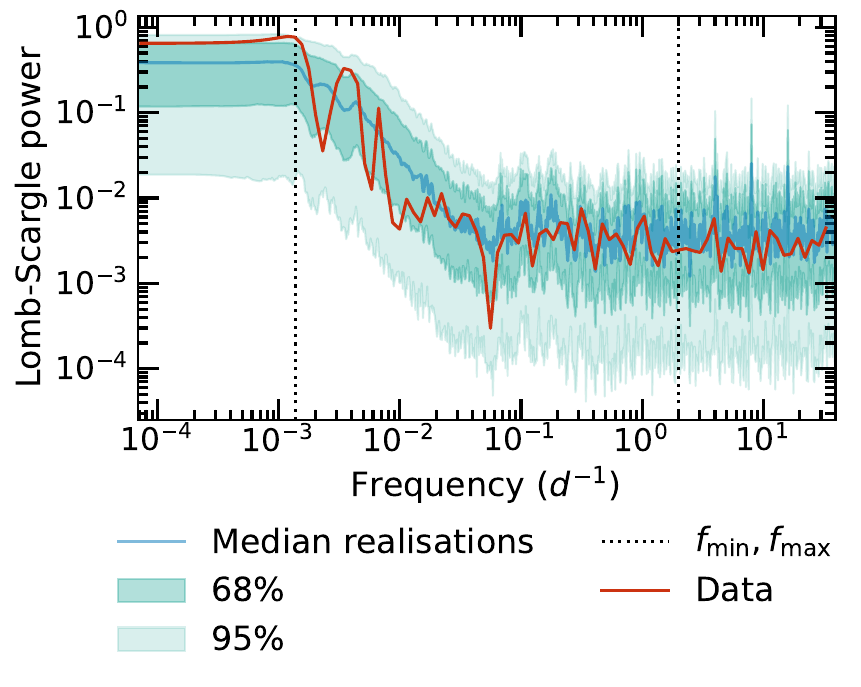}
        \caption{Posterior predictive Lomb-Scargle periodogram from realisations of the conditioned Gaussian process.}
        \label{fig:diag_ppc_lsp}
    \end{subfigure}
    \caption{Diagnostics post-inference using the posterior samples.}
    \label{fig:diagnosticsd}
\end{figure*}

To check whether the inference yielded a `good fit', we provide several visual diagnostics using samples from the posterior distribution. As an example, we use the results of one simulation and present these diagnostics in \cref{fig:diagnosticsd}.

\subsubsection{In the time domain}

Given a model and parameter values $\boldsymbol{\theta}$, we can predict the time series at any time by conditioning the Gaussian process on the observation. This assumes that we have a best-fitting value for the parameters of the models, for instance, \cite{2014ApJ...788...33K} used a maximum a posteriori estimate.

Instead of using a single point estimate, here we use a subset of the posterior samples generated with MCMC or NS. For each set of parameter values, we condition the GP on the observed data and the error bars $\boldsymbol{\sigma^2}$ using \cref{eq:conditionmean,eq:conditioncov} to obtain a conditioned GP. From this process, we then draw a realisation. After iterating over the subset of posterior samples, we have a distribution of realisations, from which we compute the median and quantiles and study how accurate the GP is at predicting the data.

This is shown in \cref{fig:diag_pred} where the shaded areas show the quantiles of the predictive distribution. The noisy variations observed at the boundaries of prediction bands are due to the finite number of samples drawn, if we had drawn more samples these boundaries would appear smoother. It should be stressed that the covariance matrix of the conditioned GP may not retain the nice quasi-separable properties of the original GP, which makes sampling and drawing realisations from the conditioned GP very slow for large datasets. \\

Using the samples from the conditioned GP we also compute the standardised residuals defined as $r = (y-y_\mathrm{sim})/\sigma_y$. If the observation is only contaminated by Gaussian white noise and if the model accurately reproduces the red noise time series, the residuals should be white, i.e. Gaussian distribution with zero-mean. In the upper panel of \cref{fig:diag_res}, we show the time series of the residuals and the distribution of residuals. We observe that the residuals resemble a Gaussian white noise. Following \cite{2014ApJ...788...33K}, we plot in the lower panel the autocorrelation function (ACF) of the residuals, if the residuals are not equispaced in time then the time lags of the ACF are not proper time lags but lags between indexes. This may explain the deviation of the second value of the ACF from the expected distribution of white noise.

A possible way to check if non-equispaced residuals follow a white noise distribution could be to fit a first-order autoregressive process to the residuals. If the autoregressive coefficient is close to zero then residuals follow a white noise.

\subsubsection{In the Fourier domain}

First, it is recommended to check that the power spectrum approximation in terms of $J$ basis functions still holds with the posterior samples. This can be checked by comparing the power spectrum model, and the approximated power spectrum model using the posterior samples as shown in \cref{fig:diag_ppc_psd}. In this case, we do not see any notable difference between the model in orange and the approximation in blue. If the approximated power spectrum model were to deviate from the intended model (bending power law), e.g. if we discern the shape of individual basis functions in the approximated model then we advise increasing the number of basis functions and restarting the inference. As the normalisation of the power spectrum is defined with the total variance, the noise level is given by $P_\mathrm{noise}=2 \nu \Delta t \overline{\sigma_\mathrm{err}^2}$ \citep[see appendix A of ][]{2003MNRAS.345.1271V}.\\

Diagnostics in the Fourier domain are critical to assess whether features were missed in the modelling. We suggest drawing realisations from the GP -- using the same sampling pattern -- with samples from the posterior distribution and computing the Lomb-Scargle periodogram \citep{1976ApjSS..39..447L,1982ApJ...263..835S}. In \cref{fig:diag_ppc_lsp}, we compare the distribution of these periodograms to the Lomb-Scargle periodogram of the observation and find a good agreement. If we find some features missing in the periodogram of the realisations, we could include them in the modelling and compare the two models. We stress that this diagnostic based on the periodogram of irregular time series in the Fourier domain is challenging and may strongly depend on the sampling pattern. Other diagnostics in the Fourier domain could involve the Fourier transform of the discrete correlation function \citep{1988ApJ...333..646E}. The purpose of these diagnostics is to ensure that the model is not completely wrong. For instance, these are not used to check whether the estimated bend(s) or slopes agree with the Lomb-Scargle periodogram which is distorted by the sampling pattern and can be noisy. This can be seen in \cref{fig:diag_ppc_lsp} where the periodogram does not show signs of a bend.

\section{Simulation-based calibration}
\label{sec:simulationbased}
To validate the method presented here, we simulate several types of fake time series and try to recover the input parameters.

\subsection{Simulating random time series}
\label{sec:simu}
We simulate a time series $\boldsymbol{x}$ given a power spectrum model following the method of \cite{1995A&A...300..707T}\cite[see also][]{Ripley87}.
To include any effect of aliasing or leakage, the time series is simulated with a longer duration and a finer sampling than what is required, and we use only a subset of this simulated time series. In the limit of an infinite number of data points, this method is equivalent to drawing realisations from a Gaussian process.

To produce a log-normal distribution of values for $\boldsymbol{y}$, the time series $\boldsymbol{x}$ is exponentiated, i.e. ${y}_i= \exp {x}_i$. For each simulated value, a measurement error is simulated as $\epsilon_i = \beta\sqrt{y_i}$ where $\beta$ is a real number drawn from the interval $[1,4]$ where the probability is such that $P(\beta=1)=0.99$ and $P(\beta=4)=0.01$. To simulate the observation process, the observed value $y_i^\mathrm{obs}$ is drawn from the Gaussian distribution: $\mathrm{Normal}(y_i,\epsilon_i^2)$.

Finally, we generate random gaps in the time sampling. The number and duration of the gaps are drawn from uniform distributions with the condition that the duration of the gappy time series should be no less than $75\%$ of the input duration. In our simulations, we set the condition that in the final time series, $F_\text{var}$ should not exceed $1.5$. In \cref{tab:simulationsinput}, we present the two sets of simulations used in this work. We simulate time series with a single-bending power-law power spectrum for short and long observations. For the short observation, we also simulate time series with a steep power spectrum, i.e. with a high-frequency slope $\alpha_2 \sim \text{Uniform}[3,6]$.

\begin{table}
    \centering
    \caption{Summary of the simulation set-up with details about the time series and the input model. \textsuperscript{\textdagger}The values of $f_\text{min}$ and $f_\text{max}$ depend on the duration and sampling of the observation.}
    \label{tab:simulationsinput}
    \begin{tabular}{lcc}\toprule\toprule
        Dataset               & short                                                                                                         & long     \\ \midrule
        Duration              & 2 years                                                                                                       & 16 years \\
        Minimum sampling      & 6 hours                                                                                                       & 6 hours  \\
        Number of points      & 500                                                                                                           & 1000     \\
        Number of simulations & 4000                                                                                                          & 4000     \\
        Number of gaps        & 0-3                                                                                                           & 3-6      \\
        \midrule
        Bending power-law     & \multicolumn{2}{c}{$\alpha_1 \sim \text{Uniform}[0,1.25]$}                                                               \\
        parameters            & \multicolumn{2}{c}{$f_1 \sim \text{Log-uniform}[4 f_\text{min},f_\text{max}/4]$\textsuperscript{\textdagger}}            \\
                              & \multicolumn{2}{c}{$\alpha_2 \sim \text{Uniform}[1.5,4]$}                                                                \\\\
        Steep model           & $\alpha_1$ and $f_1$ identical as above                                                                       & -        \\
                              & $\alpha_2 \sim \text{Uniform}[3,6]$                                                                           & -        \\
        \bottomrule\bottomrule
    \end{tabular}
\end{table}

\subsection{Priors and inference}

To infer the power spectrum of the simulated time series, we use the nested sampling (NS) and Hamiltonian Monte Carlo (HMC) codes presented in the previous section. NS is used for all datasets while HMC is only used on the short dataset. For the dataset of short time series, two types of priors for the slope $\alpha_2$ are used. We use prior distributions identical to the input of simulations shown in \cref{tab:simulationsinput}, or with a prior on $\alpha_2$ conditional on the value of $\alpha_1$, this can written as $\pi(\alpha_2|\alpha_1)$.

Prior parameters for the time series or the Gaussian process are the same as shown in \cref{tab:modelling}. We also account for the log-normal distribution of values with a logarithmic transformation as the time series were exponentiated. For all simulations, the prior on the mean is defined with a zero-mean normal distribution with a variance of 4. The prior distribution of the variance is given by a log-normal distribution with parameters $\mu=s^2$ and $\sigma^2=1$, where $s^2$ is the sample variance of the time series. In this case, the prior distribution of the variance is not exactly Bayesian as it depends on the data. In these simulations, $\mu, \nu, c$ and the variance are considered as nuisance parameters, we are interested in checking if we can recover the parameters of the power spectrum model.

\subsection{Validating the method}

\begin{figure*}
    \centering
    \includegraphics[width=\textwidth]{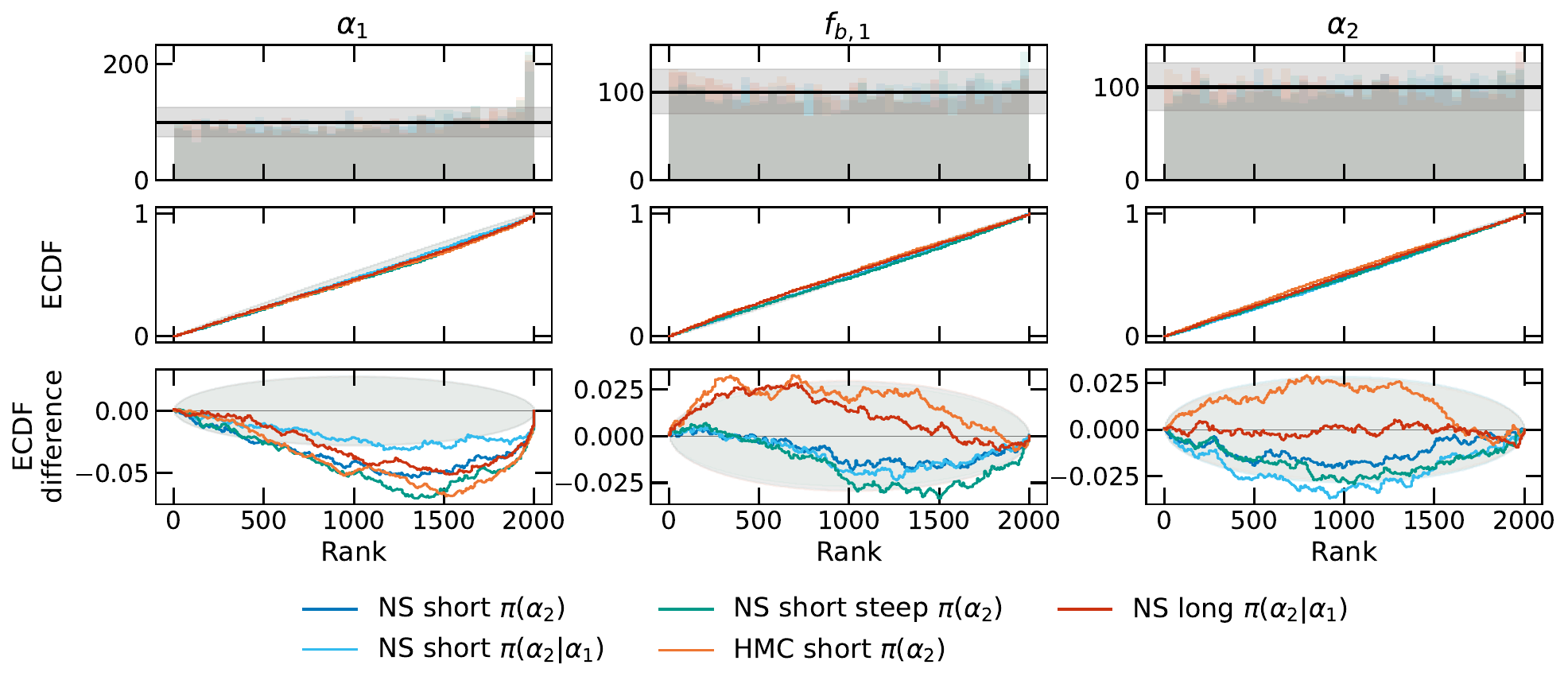}
    \caption{Simulation-based calibration. The top row shows the distribution of the rank statistic for the three parameters, and the horizontal black line shows the expected uniform distribution for the rank with a $99\%$ confidence band. The middle row shows the empirical cumulative distribution function (ECDF) and the bottom row shows the difference between the ECDF and the theoretical CDF. The band represents a $99\%$ confidence level of the theoretical CDF. NS and HMC, respectively correspond to inference with nested sampling and Hamiltonian Monte Carlo. $\pi(\alpha_2|\alpha_1)$ indicates that the prior on $\alpha_2$ is conditional on the value of $\alpha_1$. Short, short steep and long correspond to the length of the simulations presented in \cref{tab:simulationsinput}.}
    \label{fig:SBC}
\end{figure*}

Having defined a modelling framework, we now wish to establish its veracity. Computing the median or mean of the posterior distribution and comparing it to the true value may not be adequate in a Bayesian framework as they are simply point estimates.
We use simulation-based calibration \citep{cook2006,2018arXiv180406788T} to assess the quality of the method and ensure that the distribution of the posterior samples is consistent with the input distribution of the simulated datasets. This method can be summarised as follows: draw parameter samples from the prior distribution, simulate time series using the samples, collect posterior samples for all parameters and simulations, and then compute the rank statistic of the prior sample relative to the posterior sample. The rank statistic can be seen as the index of the prior sample if it were to be inserted in an ordered sample of posterior values. \cite{2018arXiv180406788T} showed that the rank statistic is expected to follow a discrete uniform distribution if the samples from the posterior distribution match the true underlying posterior.\\

In \cref{fig:SBC}, we present the graphical diagnostics introduced in \cite{2021arXiv210310522S} to assess the method. They diagnose when the distribution of the rank statistic (top row) deviates from its expected uniform distribution using the empirical cumulative distribution function (ECDF) in the middle and bottom rows.
The second and third columns show that $f_{b,1}$ and $\alpha_2$ follow their expected distribution. We find that $\alpha_1$ may be overestimated, however, if we restrict the simulations to larger values of $f_{b,1}$ this bias vanishes. This means that when the bend is close to the minimum observed frequency, the estimation of the low-frequency slope is difficult and can be biased. We find however that the posterior distributions are not over- or under-dispersed, meaning there is no over-fitting. Inference with NS and HMC sampling algorithms agrees in most cases.

\section{Application to Ark 564}
\label{sec:ark564}

We demonstrate how this method can model the power spectrum of X-ray fluctuations in the active galaxy Ark 564. We use observations from the \textit{Neil Gehrels Swift Observatory} \citep{2004ApJ...611.1005G} and \textit{XMM--Newton} \citep{2001A&A...365L...1J} spanning nearly 22 years.

\subsection{Data reduction and calibration}

\subsubsection{Swift-XRT}
\begin{figure*}
    \centering
    \includegraphics[width=\textwidth]{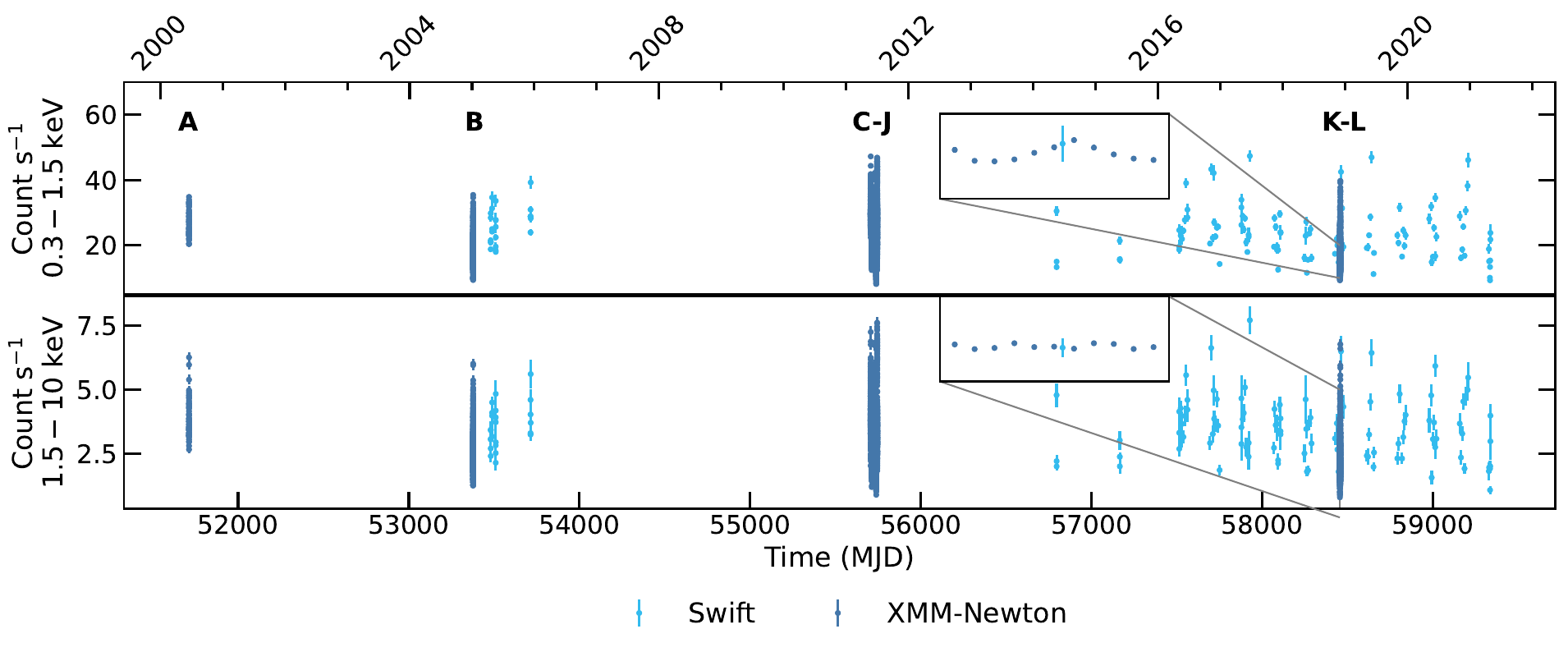}
    \caption{Combined \textit{Swift} and \textit{XMM--Newton} light curves in the soft (top) and hard (bottom) bands. The insets show the simultaneous period between \textit{XMM--Newton} and \textit{Swift}. The letters refer to the \textit{XMM--Newton} observations shown in \cref{fig:xmmlc} and presented in \cref{tab:ark564xmm}.}
    \label{fig:xmmlc_simultaneous}
\end{figure*}

We extract X-ray light curves from the \textit{X-ray telescope} \citep{2005SSRv..120..165B} on board \textit{Swift} with the online tool\footnote{\url{https://www.swift.ac.uk/user_objects/}} described in \cite{2007A&A...469..379E,2009MNRAS.397.1177E}. The light curves are binned by snapshots, which means that we have one point per orbit of the spacecraft. The light curves are corrected for pile-up, bad columns on the CCD and vignetting.  We extract the light curves in the soft ($0.3-1.5$~keV) and hard ($1.5-10$~keV) energy bands. We only keep data obtained in the Photon Counting mode (PC). \cref{fig:xmmlc_simultaneous} shows the \textit{Swift-XRT} light curve (light blue).

\subsubsection{XMM--Newton}

\begin{figure*}
    \centering
    \includegraphics[width=\textwidth]{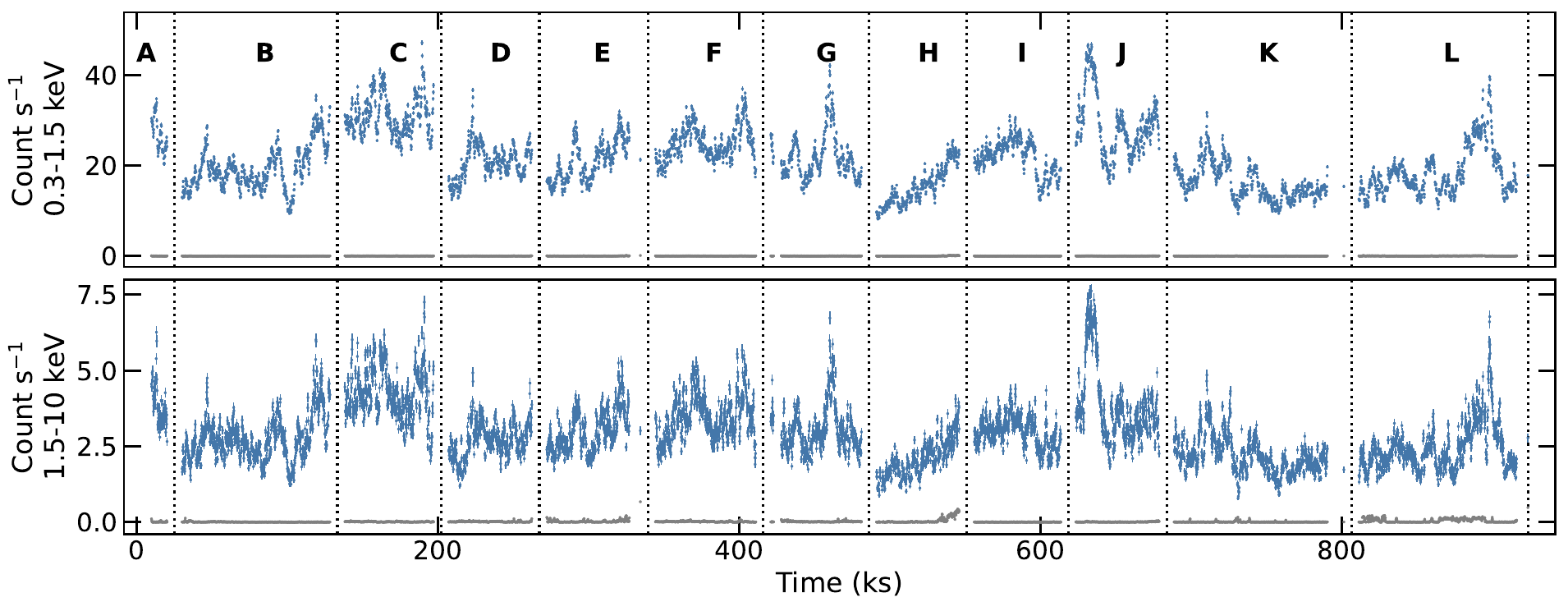}
    \caption{Concatenated \textit{XMM--Newton} light curves in the soft (top) and hard (bottom) energy bands. The background rate is shown in grey. The letters refer to the \textit{XMM--Newton} observations presented in \cref{tab:ark564xmm}.}
    \label{fig:xmmlc}
\end{figure*}

We select twelve \textit{XMM--Newton} observations from 2000 to 2018 listed in \cref{tab:ark564xmm}. Data from the EPIC-pn camera \citep{2001A&A...365L..18S} were reduced using the Science Analysis System (SAS) 21.0 \citep{2004ASPC..314..759G}. All these observations were performed in small window mode, we selected the source in a circular region of radius $25$ of arcseconds. The background was selected in a source-free circular region of $50$ arcseconds radius. First, event lists for the source and background regions are extracted with \texttt{PATTERN<=4} to include single and double events. Then, counts are binned in light curves with a bin time of $\Delta t =150$\,s. Finally, light curves are corrected for bad time intervals and loss of exposure\footnote{Bins containing one or several bad time intervals have their exposure reduced, to correct for this effect the number of counts in the bin is rescaled proportionally to the lost exposure. Bins with zero exposure are removed.}. No pile-up was noticeable in the observations. The  \textit{XMM--Newton} light curves in the soft and hard bands are shown in \cref{fig:xmmlc} (dark blue).

\begin{table}
    \centering
    \caption{\textit{XMM--Newton} observations used in this work with the net exposure time and the number of points in the extracted light curve with $\Delta t= 150$\,s. The light curves of all epochs are plotted in \cref{fig:xmmlc}.}
    \begin{tabular}{lllll}
        \toprule\toprule
        Epoch & Observation & Date         & Net exposure (ks) & N   \\\midrule
        A     & 0006810101  & 2000 June 17 & 10.3              & 70  \\
        B     & 0206400101  & 2005 Jan 5   & 98.6              & 654 \\
        C     & 0670130201  & 2011 May 24  & 58.8              & 393 \\
        D     & 0670130301  & 2011 May 30  & 55.2              & 369 \\
        E     & 0670130401  & 2011 June 5  & 54.8              & 366 \\
        F     & 0670130501  & 2011 June 11 & 66.6              & 445 \\
        G     & 0670130601  & 2011 June 17 & 60.2              & 368 \\
        H     & 0670130701  & 2011 June 25 & 54.9              & 367 \\
        I     & 0670130801  & 2011 June 29 & 57.5              & 384 \\
        J     & 0670130901  & 2011 July 1  & 55.2              & 368 \\
        K     & 0830540101  & 2018 Dec 11  & 101.7             & 679 \\
        L     & 0830540201  & 2018 Dec 3   & 104.7             & 695 \\ \bottomrule\bottomrule
    \end{tabular}
    \label{tab:ark564xmm}
\end{table}

\subsubsection{Inter-calibration}
To combine the two light curves, we compare the count-rate for both instruments during a simultaneous observation in 2018 as shown in the inset of \cref{fig:xmmlc_simultaneous}. A scaling factor $\gamma$ is applied on the \textit{Swift} data to match the \textit{XMM--Newton} count-rate. The scaling factor is obtained by averaging the two \textit{XMM--Newton} data points enclosing the \textit{Swift} data. We find that the \textit{Swift} count-rate must be multiplied by a factor $12.0$ and $6.56$ for the soft and hard bands respectively.

\begin{figure}
    \centering
    \includegraphics[width=.5\textwidth]{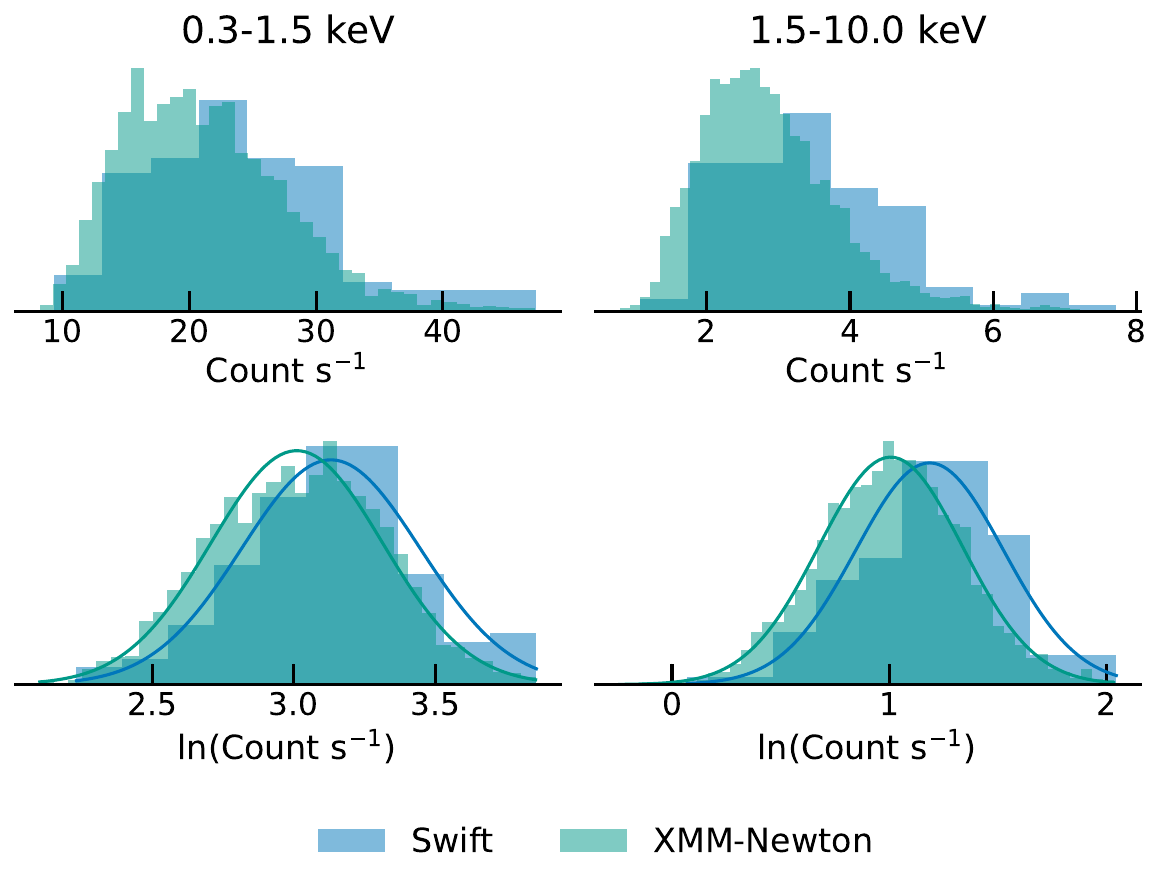}
    \caption{Count-rate distribution of \textit{XMM--Newton} and \textit{Swift} light curves in the soft (left) and hard (right) energy bands (top panels). Distribution of the logarithm of the count-rates fitted with a normal distribution (bottom panels).}
    \label{fig:fluxdistrib}
\end{figure}

In \cref{fig:fluxdistrib}, the distributions of the count-rates are shown, and the bottom panels show the distribution of the logarithm of these values. The distribution of the log of the count-rates appears to be well-modelled with a normal distribution. The distributions for both instruments overlap but do not perfectly match, an additional parameter will be added to account for the calibration.

\subsection{Modelling}

We infer the shape and parameters of the single and double-bending power-law model. As introduced in \cref{sec:approx}, we extend the low and high frequencies with the scale factors $S_{\rm low}=S_{\rm high}=20$. We use $J=30$ basis functions. The priors on the power spectrum parameters were presented in \cref{tab:modelling}. We apply a logarithmic transformation to the data, as presented in \cref{sec:logtransform}. We tried three different distributions for the prior on the shift of the logarithm transformation. A log-uniform distribution, a Half-Cauchy distribution and a log-Half-Cauchy distribution, all yielded identical posterior distributions for all other parameters with similar Bayesian evidence. We present the results with the log-Half-Cauchy prior in \cref{tab:results_Ark564}.

To estimate the mean and variance of the process, we use the priors defined previously in \cref{sec:priorsGP}. We extract three per cent of the original time series to compute the initial values for the prior on the mean, these values are presented in \cref{tab:ark564_meanvar}. We account for the log-normal distribution of count rates using the transformation presented in \cref{sec:logtransform}. As we find that the high-frequency slope $\alpha_3$  may be steeper than $4$, we also use the steep basis function $\psi_6$ for the approximation.

\begin{table}
    \centering
    \caption{Initial values for the Gaussian prior on the mean for the log-transformed soft and hard energy bands.}
    \begin{tabular}{lll}\toprule
                                   & $0.3-1.5$ keV & $1.5-10$ keV \\ \midrule
        $\bar{x}$ (Count s$^{-1}$) & $3.01$        & $1.01$       \\
        $s^2$ (Count$^2$ s$^{-2}$) & $0.0920$      & $0.117$      \\\bottomrule
    \end{tabular}
    \label{tab:ark564_meanvar}
\end{table}

\subsection{Results and discussion}

\subsubsection{Results}

The posterior samples for all the models and energy bands are given in \cref{apdx:posteriors}. Posterior medians and uncertainties from the $16^\text{th}$ and $84^\text{th}$ percentiles are given in \cref{tab:results_Ark564}. The last row shows the logarithm of the Bayes factor against the single-bending power-law model.
We find very strong evidence for two bends in the power spectrum of Ark 564 as the Bayes factors are all very high ($BF\gg10^2$).

The \textit{XMM--Newton}/\textit{Swift} inter-calibration factor $\gamma$ is close to unity which shows that empirical inter-calibration using the simultaneous segment is correct. We find that the factor $\nu$ for uncertainties on measurements is less than unity. This signifies that the uncertainties are overestimated. The noise level in the power spectrum defined as $2\Delta t \nu \overline{\sigma_\mathrm{err}^2}$ is then decreased.

\begin{figure*}
    \centering
    \includegraphics[width=\textwidth]{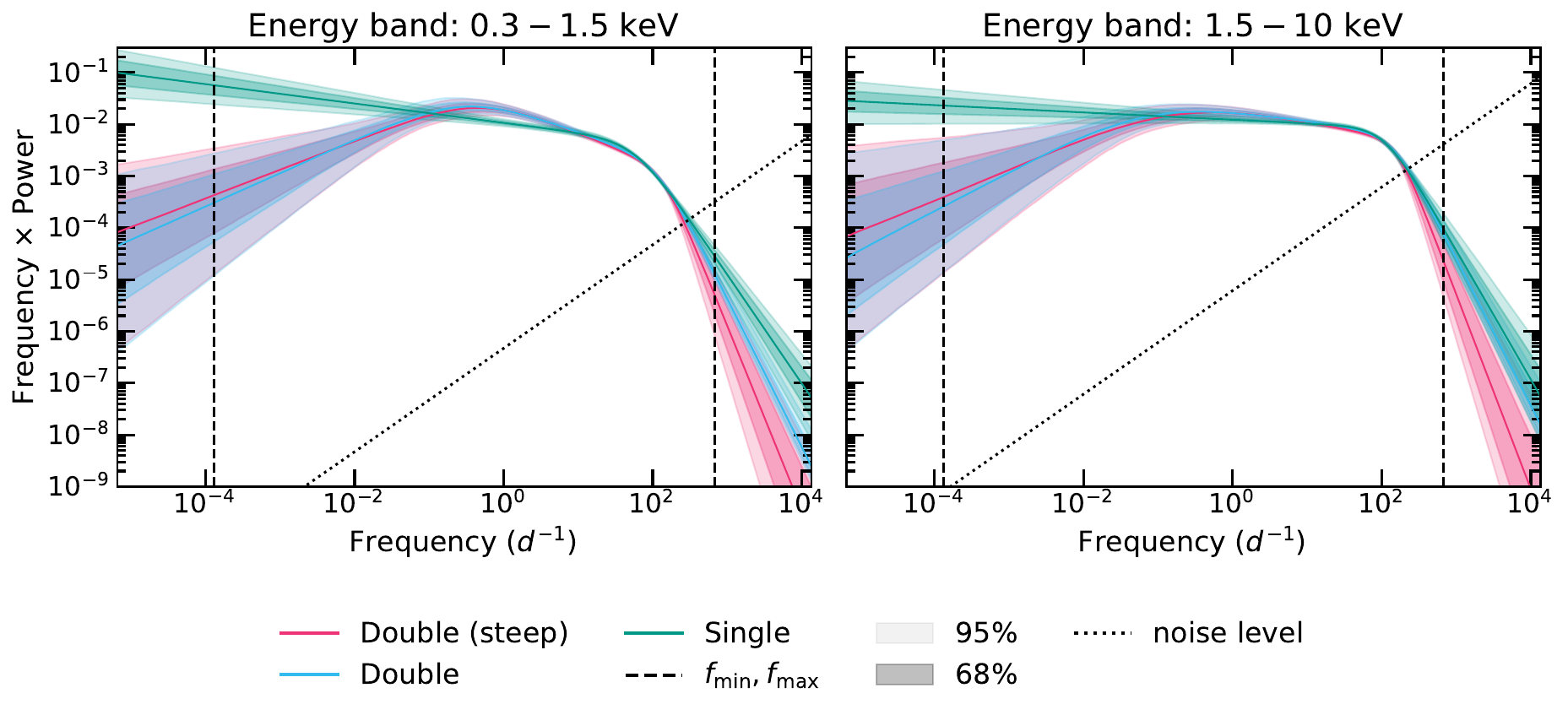}
    \caption{Posterior predictive posterior power spectra for the soft (left) and hard band (right). The single-bending power-law is plotted in green, the double-bending power-law is shown in blue and the double-bending power-law with steep basis functions is shown in pink. The dark shaded areas represent the $68\%$ confidence while the light areas represent the $95\%$ confidence.}
    \label{fig:psd_ark564_full}
\end{figure*}

The posterior predictive power spectra for the three models are shown in \cref{fig:psd_ark564_full} for the soft and hard bands. The two bends are distinguishable and we observe the difference between the approximations using $\psi_4$ and $\psi_6$. \cref{fig:posteriors_Ark564} shows the posterior distributions for all the parameters of the double-bending model approximated using $\psi_6$, the prior distributions are shown with dotted lines.
The median value on $c$ the shift in the logarithm transformation is very small, this is due to the posterior distribution being nearly equal to the prior distribution as observed in the Figure.

\renewcommand{\arraystretch}{1.5}\begin{table*}
    \centering
    \caption{Results of the power spectrum estimation of Ark 564. Median values of the posterior distributions with the $16^{\rm th}$ and  $84^{\rm th}$ percentiles. The evidence $\ln Z$ and the logarithm of the Bayes factor comparing against the single-bending power-law model.}\label{tab:results_Ark564}
    \begin{tabular}{*{7}{l}}\toprule\toprule                             & \multicolumn{3}{c}{$0.3-1.5$ keV} & \multicolumn{3}{c}{$1.5-10$ keV}                                                                                                                     \\
               Parameter                                             & Single                            & Double                           & Double (steep)             & Single                     & Double                     & Double (steep)             \\ \midrule$\alpha_1$ & $1.19\pm{0.04}$ & $0.36^{+0.21}_{-0.30}$ & $0.44^{+0.18}_{-0.28}$ & $1.07\pm{0.03}$ & $0.25^{+0.33}_{-0.32}$ & $0.41^{+0.27}_{-0.37}$ \\
               $f_{b,1}$ (d$^{-1}$)                                  & $58.86^{+10.10}_{-10.12}$         & $0.31^{+0.31}_{-0.15}$           & $0.50^{+0.50}_{-0.25}$     & $110.37^{+12.45}_{-13.24}$ & $0.10^{+0.25}_{-0.06}$     & $0.25^{+1.13}_{-0.17}$     \\
               $\alpha_2$                                            & $3.11^{+0.20}_{-0.17}$            & $1.53\pm{0.06}$                  & $1.60\pm{0.07}$            & $3.49^{+0.28}_{-0.29}$     & $1.22^{+0.06}_{-0.05}$     & $1.29^{+0.10}_{-0.07}$     \\
               $f_{b,2}$ (d$^{-1}$)                                  & -                                 & $116.87^{+11.92}_{-10.98}$       & $138.45^{+12.14}_{-13.34}$ & -                          & $139.09^{+13.78}_{-13.10}$ & $158.73^{+14.25}_{-12.53}$ \\
               $\alpha_3$                                            & -                                 & $3.87^{+0.10}_{-0.17}$           & $4.55^{+0.59}_{-0.47}$     & -                          & $3.84^{+0.12}_{-0.22}$     & $4.78^{+0.66}_{-0.61}$     \\
               \midrule variance (Count$^2$ s$^{-2}$)                & $0.58^{+0.30}_{-0.19}$            & $0.11^{+0.02}_{-0.01}$           & $0.11^{+0.02}_{-0.01}$     & $0.28^{+0.09}_{-0.07}$     & $0.12^{+0.02}_{-0.01}$     & $0.12^{+0.02}_{-0.01}$     \\
               $\nu$                                                 & $0.41\pm{0.14}$                   & $0.54^{+0.10}_{-0.11}$           & $0.75^{+0.15}_{-0.16}$     & $0.71\pm{0.10}$            & $0.70^{+0.09}_{-0.10}$     & $0.86^{+0.11}_{-0.12}$     \\
               $\mu$ (Count s$^{-1}$)                                & $3.09\pm{0.50}$                   & $3.07\pm{0.07}$                  & $3.06\pm{0.07}$            & $1.04^{+0.29}_{-0.31}$     & $1.05\pm{0.08}$            & $1.05\pm{0.08}$            \\
               $\ln c$ (Count s$^{-1}$)                              & $-23.13^{+9.26}_{-20.47}$         & $-23.34^{+9.40}_{-18.56}$        & $-23.22^{+9.10}_{-19.26}$  & $-23.92^{+9.19}_{-19.33}$  & $-23.39^{+8.71}_{-19.41}$  & $-23.64^{+9.20}_{-19.58}$  \\
               $\gamma$                                              & $0.94\pm{0.07}$                   & $0.93\pm{0.06}$                  & $0.92^{+0.07}_{-0.06}$     & $0.83^{+0.10}_{-0.09}$     & $0.86\pm{0.07}$            & $0.85\pm{0.07}$            \\
               \midrule$\ln Z$                                       & $7639.12$                         & $7663.33$                        & $7665.48$                  & $3858.01$                  & $3865.96$                  & $3869.99$                  \\
               Log Bayes factor $\ln BF=\ln Z-\ln Z_\mathrm{single}$ & -                                 & $24.21$                          & $26.36$                    & -                          & $7.95$                     & $11.98$                    \\
               \bottomrule\bottomrule\end{tabular}\end{table*}

\subsubsection{Comparison with previous works}

Using \textit{RXTE}, \textit{ASCA} and \textit{XMM--Newton} observations \cite{2007MNRAS.382..985M} found that the periodogram of Ark 564 was well modelled with either a double-bending power-law model or with two Lorentzians. The energy bands are slightly different from ours: $0.6-2$~keV for the soft band and $2-8.8$~keV for the hard band. The low-frequency slope was fixed to zero in their modelling.

\begin{figure}
    \centering
    \includegraphics[width=.5\textwidth]{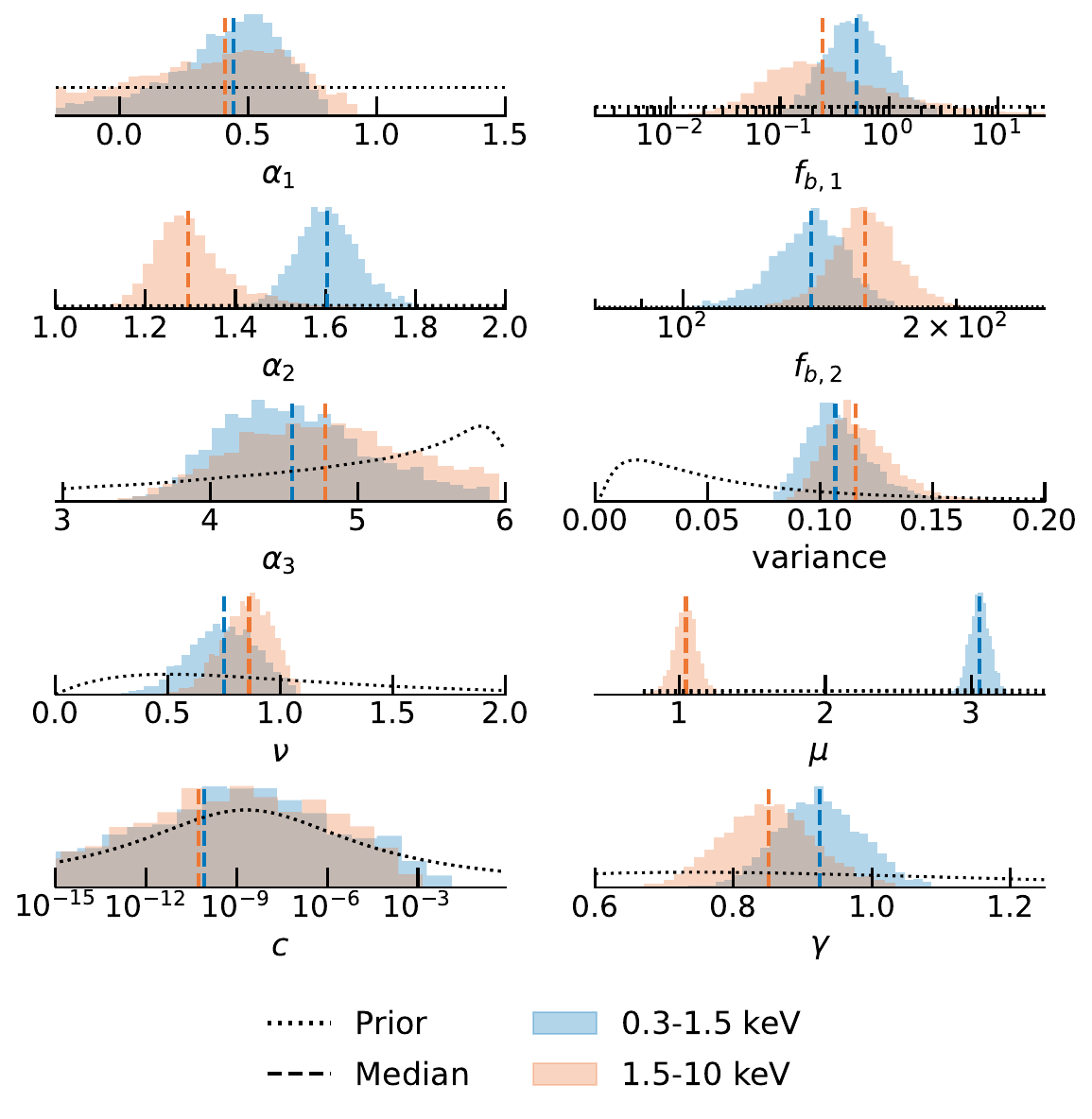}
    \caption{Distribution of the posterior samples for the double-bending power-law approximated with $\psi_6$, in the soft band (blue) and hard band (orange). The medians are shown with vertical dashed-dotted lines. The priors are shown in black dashed lines.}
    \label{fig:posteriors_Ark564}
\end{figure}

We find that all our estimates of the bends and slopes strongly agree with the results previously obtained. This was done using nearly independent data -- only the 2005 \textit{XMM--Newton} observation is used in both works. \cref{fig:posteriors_Ark564}
shows that the low-frequency slope $\alpha_1$ is consistent with zero and that the intermediate slope is steeper for the soft band. The high-frequency slope is steeper than 3.5 with a bend corresponding to a time-scale of about 10 minutes.
It is difficult to distinguish differences in the posterior distributions of $f_1$, $f_2$ and $\alpha_3$ between the soft and hard bands.

\section{Discussion}
\label{sec:discussion}

Previous work has shown that the variability of AGN (and also often other accreting compact objects such as X-ray binaries and cataclysmic variables)  is dominated by broad-band noise with a power spectrum that can be described reasonably well with power-laws connected by smooth bends. The bend frequencies represent characteristic time-scales of the variable system and must be related to the size, geometry and physics of the accretion flow and radiative mechanisms. Here, we have developed a new method, \texttt{PIORAN}, specifically for fitting this form of power spectral model and parameter estimates, using a fast Gaussian process approach which avoids the biases and lack of calculable likelihood function associated with previous Fourier-based forward-modelling approaches. We now compare our method with other Gaussian process approaches for fitting power spectra without the constraints present in e.g. DRW approach, before discussing areas where \texttt{PIORAN} might be improved.

\subsection{Standard Gaussian process methods}
\label{sec:fftmethod}

As introduced in \cref{sec:gp}, Gaussian processes allow modelling of power spectral densities, where the functional form of the covariance constrains the shape of the power spectrum. \cite{2010MNRAS.403..196M} and \cite{2013ApJ...777...24Z} proposed to specify a power spectrum model, apply the discrete Fourier transform to compute numerically the autocovariance function and then use it in the Gaussian process regression.

This method suffers from two drawbacks, first, it cannot be applied to large datasets ($N\gtrsim 10^3$) as it uses the standard Gaussian process approach which is computationally expensive (see \cref{sec:likelihood}). Secondly, with some tests, we find that when the grid of frequencies is large ($\gtrsim 10^6$), the fast Fourier transform becomes a bottleneck. This can arise when a time series contains values separated by long time-scales and short time-scales, which can be the case for long-term light curves from \textit{Swift} and \textit{RXTE}.

In comparison, our method makes use of bending power-law power spectral model built using computationally efficient power-law-like basis functions. Thanks to the \texttt{celerite} algorithm \citep{2017AJ....154..220F}, our method is faster by at least an order of magnitude (see \cref{fig:cost}) and can be applied to much more data points -- up to $N\simeq 50,000$ in practice.

In \cite{2011ApJ...730...52K}, the power spectrum is modelled with a weighted sum of Lorentzian basis functions, $\psi_2$. The basis functions are also geometrically spaced in frequency, the important difference with our method relies on the weights. In \cite{2011ApJ...730...52K}, the weights allow modelling of a bending power-law power spectrum described as follows: flat at low frequencies, bends to $f^{-\alpha}$ at an intermediate frequency and then bends to $f^{-2}$ at high frequencies. This weighting scheme allows estimation of the intermediate slope $\alpha$ and the two bend frequencies. Here, we allow for arbitrary slopes at low, intermediate and high frequencies as long as they agree on a decreasing power spectrum. Furthermore, our modelled power spectra can be steeper as the basis functions $\psi_4$ and $\psi_6$ are steeper than $\psi_2$.

CARMA processes of order $(p,q)$ introduced in \cref{sec:carma} allow flexible modelling of the broad-band variability in the time domain using a Stochastic Differential Equation (SDE). However, the interpretation of the coefficients of the SDE can be difficult and the choice of $p$ and $q$ is non-trivial as the models are not nested. Some simpler models that can be expressed with a few parameters (e.g. bending power-law) may need very high-order CARMA representation, with complex relations between coefficients. In practice, this modelling makes the parametrisation of the priors very difficult. Sampling from the posterior can be challenging as it will be multimodal when $p$ and $q$ are high.

In comparison, our method allows a more constrained shape for the broad-band power spectrum (bending power-laws) and a simpler parametrisation using only the parameters of the bending power-law model. CARMA processes can model narrow features in the power spectrum, with our method it could be possible to add such features manually after the bending power-law model is approximated.

\subsection{\texttt{PIORAN}}

Like all Gaussian process regression methods, \texttt{PIORAN} assumes Gaussian data and sampled values in the time series. While this method can be applied to most AGN time series, here, we describe ideas on how one could lift some of these assumptions.

\subsubsection{Stationarity}

We assumed the underlying process generating the time series to be stationary. In practice, this may not be true for accreting systems. For instance, it is well-known that the energy spectrum of X-ray binaries can drastically transition from soft to hard state \citep{2006csxs.book..157M}. The physical processes dominant for the emission of photons are not the same between these states.
For possibly different reasons, a similar behaviour is observed in the class of \textit{Changing-look} AGN where the AGN transitioned from one class to another \citep{2023NatAs...7.1282R}.

A way to study non-stationary signals could be to infer the power spectrum of contiguous segments of the time series and check whether they agree on the same model. However, choosing the number and duration of the segments might be challenging without any spectral information. A more general approach could be to model the non-stationarity with deep-state Gaussian processes \citep{2020arXiv200804733Z}.

\subsubsection{Gaussianity and sampling}

In this work, we assume log-normal fluxes and apply a logarithmic transformation to the data to make the data Gaussian. As we use Gaussian processes, outliers are expected to be very rare, this means that tidal disruption events (TDEs), outbursts or flaring events cannot be well-modelled. A simple solution could be to filter these events from the time series by visual inspection of the posterior predictive time series after inference. A distribution with a broader tail might be more robust to outliers, for instance a Student-t process \citep[e.g.][]{2018arXiv180106147T}.

Another process to consider is the photon-counting nature of X-ray counts, which can be modelled with a Poisson distribution. Assuming the background is negligible, the time series is then Poisson-distributed. Using a hierarchical Bayesian modelling one could account for Poisson  data with the approach presented in this paper. However, this may be computationally expensive as each value of the time series will become a parameter of the hierarchical model. In practice, one can use approximate Bayesian methods such as variational inference to infer credible values for the process.

One should note that even periodogram-based methods also assume sampled values rather than binned values. It is be possible to account for binned values in the Gaussian process framework with uncertain inputs. However, similarly to Poisson distributions, this requires a hierarchical Bayesian model which is be computationally expensive.

\subsubsection{Uncorrelated noise}

In our modelling of the time series, we assumed the measurement variances to be independent and identically distributed. This is modelled by adding the term $\boldsymbol{\sigma^2}I$ to the diagonal of the covariance matrix. In the case of time series from binned X-ray counts, the errors are uncorrelated but other time series may not have this property. For instance, it is critical to model the observation noise for exoplanet detection. Correlated noise can be modelled with a more complex model, see for instance the S+LEAF model in \citet{2020A&A...638A..95D} which allows banded noise models.

\subsubsection{Quasi-periodic oscillations}

The models used in the present paper do not include quasi-periodic oscillations (QPOs), but can easily be extended to allow for QPOs or other power spectral features. The bending power-law power spectrum would still be approximated using basis functions and one or more additional basis functions would model narrow QPO features. For strictly periodic variability, one could add a sinusoidal autocovariance function (which corresponds to a Dirac delta in the Fourier domain) or a periodic mean function to the GP model in the time domain. These additional features need to be added after approximating the continuum power spectrum model to avoid breaking the approximation. Then, Bayes factors \citep[e.g.][]{vysakhsthesis}, or posterior predictive p-values \citep[][]{2002ApJ...571..545P,2025MNRAS.tmp..186G} can be used for model comparison (aperiodic noise with and without a QPO or strict period).

\section{Conclusions}
\label{sec:conclusion}
In this work, we presented a novel method to infer the parameters of bending power-law power spectra models for arbitrarily sampled time series. This method relies on Gaussian process regression which has no assumption about the time sampling pattern and avoids leakage and aliasing biases of the Fourier methods. It makes full use of data, time, value and error and allows for heteroscedastic data. The method approximates the power spectrum model as a sum of $J$ power-law like basis functions. With this approximation, the power spectra models cannot be steeper than $f^{-6}$, but we believe it should be steep enough for time series of accreting compact objects. Our method relies on the fast and reliable algorithm presented in \cite{2017AJ....154..220F} which enables computation of the log-likelihood with a linear scaling with respect to the number of data points in the time series. The workflow can be summarised as follows:
\begin{enumerate}
    \item Select a bending power-law model with one or two bends, see \cref{sec:bendingpowerlaw}.
    \item Define the number of basis functions ($J$), the minimum and maximum frequencies for the approximation, see \cref{sec:approx}.
    \item If the time series data are not Gaussian, apply a transformation to make the data Gaussian, here we use a logarithmic transformation see \cref{sec:logtransform}.
    \item Chose suitable priors on the parameters of the power spectrum and time series, see \cref{sec:priors}.
    \item Check the approximation by drawing samples from the prior and comparing with the selected model, see \cref{sec:checks}. If the number of basis functions is insufficient or if the priors are inadequate go back to (ii) and (iv).
    \item Sample the posterior distribution using Hamiltonian Monte Carlo or Nested sampling methods, see \cref{sec:inference}.
    \item Perform inference and check for convergence \cref{sec:inference}
    \item Assess quality of the posterior samples with respect to the priors. See first paragraph of \cref{sec:diagnostics}.
    \item Check for misspecification of the model using the diagnostics presented in \cref{sec:diagnostics}. If the model is incomplete go back to (i) to change the model.
    \item (Optional) Simulate time series from the posterior samples and use simulation-based calibration to ensure correctness of the workflow as presented in \cref{sec:simulationbased}.
\end{enumerate}

This workflow is fully Bayesian and applies to time series with arbitrary time sampling. Through extensive simulations, we checked that our method recovers the bending frequency and high-frequency slope unbiasedly with credible posterior distributions. The low-frequency slope can be overestimated if the bending frequency is close to the minimum observed frequency of the time series.

Finally, we apply this method to long-term  X-ray light curves of \textit{XMM--Newton} and \textit{Swift} observations of Ark 564. In agreement with \cite{2007MNRAS.382..985M}, we find that the power spectrum is consistent with a double-bending power-law model. Our results strongly agree with previous analyses using observations from \textit{RXTE}, \textit{ASCA} and \textit{XMM--Newton}.

In a forthcoming paper, we will estimate the long-term X-ray power spectrum of a sample of unobscured active galaxies. The method presented here could have applications beyond X-ray astronomy, it could be used for future surveys such as Vera Rubin/LSST  \citep{2019ApJ...873..111I}. Finally, this method could be extended to estimate delays and power spectra from multivariate time series as in \cite{2013ApJ...765..106Z} and \cite{2019MNRAS.489.1957W}.


\section*{Acknowledgements}
We thank the anonymous referee for their comments which improved the clarity of the text. ML acknowledges support from STFC studentships. ML thanks fruitful discussions about Fourier transforms and periodograms with Hervé Carfantan. ML thanks the University of Amsterdam and SRON for the warm welcome during the 4-week visit where part of this work was developed. The research leading to these results has received funding from the European Union’s Horizon 2020 Programme under the AHEAD2020 project (grant agreement n. 871158). This research used the ALICE High-Performance Computing Facility at the University of Leicester. This work has made use of observations obtained with \xmm, an ESA science mission with instruments and contributions directly funded by ESA Member States and NASA. This work also made use of observations and data supplied by UK \swift\ Science Data Centre and has made use of data and/or software provided by the High Energy Astrophysics Science Archive Research Center (HEASARC), which is a service of the Astrophysics Science Division at NASA/GSFC.

\textit{Julia Packages}: AdvancedHMC.jl \citep{xu2020advancedhmc}, MPI.jl \citep{Byrne2021}, Julia \citep{Julia-2017}, Turing.jl \citep{ge2018turing}.

\textit{Python libraries}: arviz \citep{arviz_2019},  corner \citep{2016JOSS....1...24F}, jax \citep{jax2018github}, matplotlib \citep{Hunter:2007}, numpy \citep{harris2020array}, scipy \citep{2020SciPy-NMeth}, ultranest \citep{2021arXiv210109675B}.
\section*{Data Availability}
The data used in this paper are publicly available to access and download from the XMM–Newton Science Archive and the UK Swift Science Data Centre. Final data products from this study can be provided on reasonable request to the corresponding author.




\bibliographystyle{mnras}
\bibliography{references} 




\appendix

\section{Fourier transforms}
\label{apdx:fourier}
The basis function $\psi_6(f)$ can be expressed as \cref{eq:decomppsi6}, using partial fraction decomposition. The first term is the DRW basis function $\psi_2(f)$ given in \cref{tab:covgp} with a known Fourier transform $\phi_2(\tau)$.
\begin{equation}
    \psi_6(f) = \frac{1}{1+f^6} = \dfrac{1}{3}\left[\dfrac{1}{1+f^2}+\dfrac{2-f^2}{1-f^2+f^4}\right]
    \label{eq:decomppsi6}
\end{equation}

The Fourier transform of the second term can be computed using Cauchy's residue theorem. The Fourier transform is given by \cref{eq:rightterm}.

\begin{equation}
    \mathcal{F}\left[\dfrac{2-f^2}{1-f^2+f^4}\right](\tau) =\pi \mathrm{e}^{-\pi |\tau|}\left(\cos(\sqrt{3}\pi |\tau|)+\sqrt{3}\sin(\sqrt{3}\pi |\tau|)\right)
    \label{eq:rightterm}
\end{equation}

Thus $\phi_6(\tau)$ is given by \cref{eq:psi6fin}.

\begin{equation}
    \phi_6(\tau) =\frac{\pi}{3}\left[\mathrm{e}^{-2\pi |\tau|}+ \mathrm{e}^{-\pi |\tau|}\left(\cos(\sqrt{3}\pi |\tau|)+\sqrt{3}\sin(\sqrt{3}\pi |\tau|)\right)\right]
    \label{eq:psi6fin}
\end{equation}

\section{Qualifying the approximation}

\subsection{Condition number}

\label{apdx:condition}
\begin{figure}
    \centering
    \includegraphics[width=0.5\textwidth]{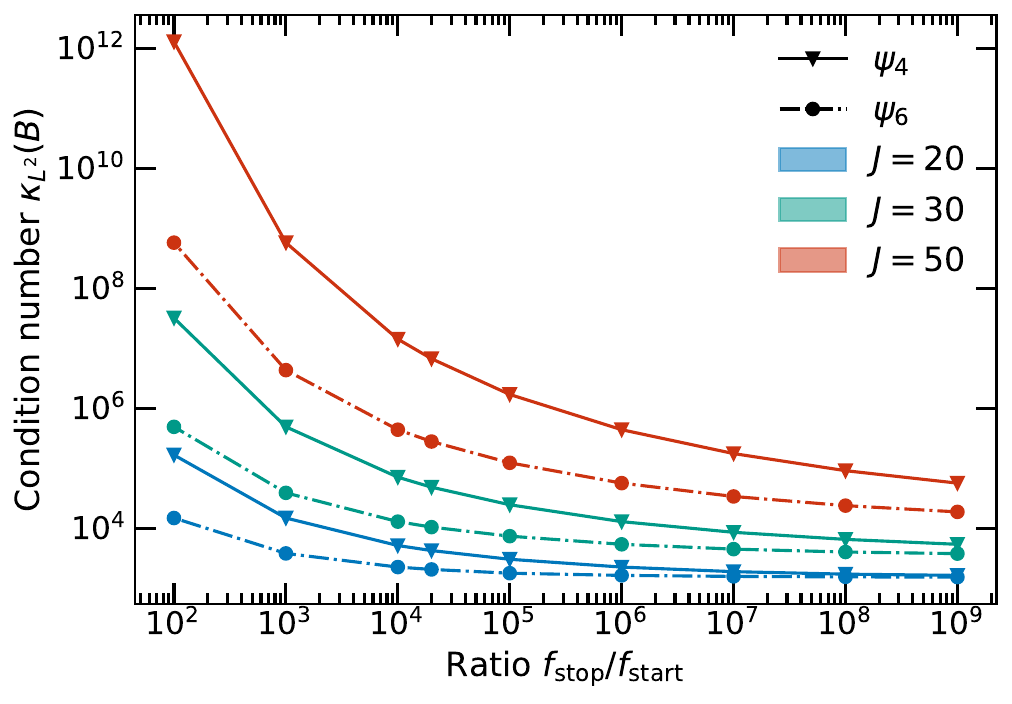}
    \caption{Condition number of the linear system for the approximation as a function of the ratio between the maximal and minimal frequency. The basis functions $\psi_4$ and $\psi_6$ are plotted respectively with triangular markers on solid lines and circle markers on dashed-dotted lines. Colours present the number of basis functions for the approximation.}
    \label{fig:conditionnumber}

\end{figure}

The condition number -- using the $L^2$ norm --  associated with the linear system of \cref{eq:approx_system} is computed as the ratio of the maximum and minimum eigenvalues. The condition number is plotted in \cref{fig:conditionnumber} as a function of the size of the frequency grid for various values of $J$. We observe that it increases with the number of basis functions used and decreases when the frequency grid spans several orders of magnitudes.

\subsection{Quality of the approximation}
\label{apdx:quality}
\begin{figure}
    \centering
    \includegraphics[width=0.5\textwidth]{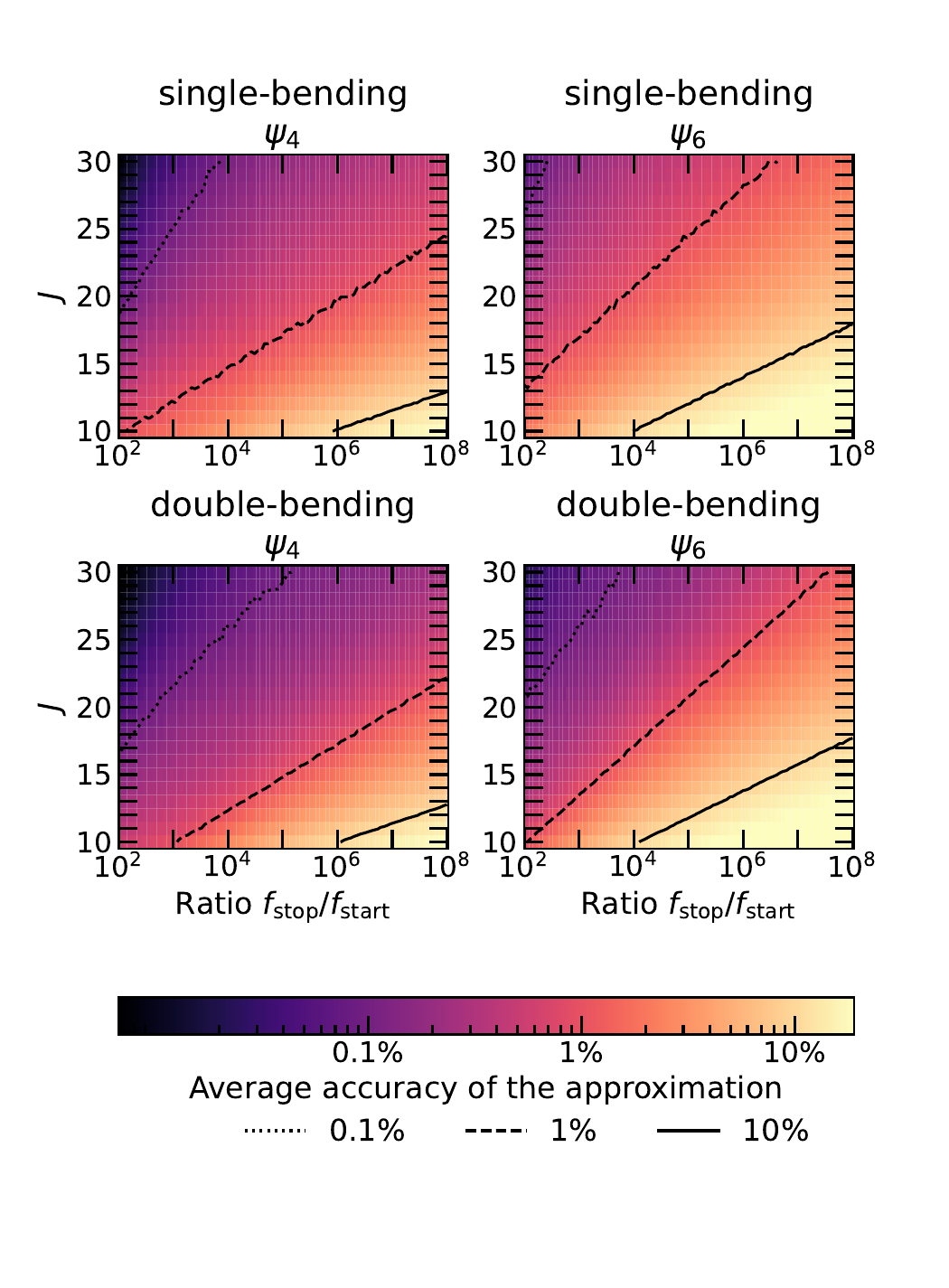}
    \caption{Average accuracy of the power spectrum approximation for $\psi_4$ and $\psi_6$ as a function $J$ (number of basis functions) and size of the frequency grid. The top panels show the accuracy for the single-bending power-law model and the bottom panel show the accuracy for the double-bending power-law model. The accuracy increases for darker colours, the dotted, dashed and solid lines represents respectively the level of $0.1\%$, $1\%$, and $10\%$ accuracy.}
    \label{fig:accuracy}
\end{figure}

The quality of the power spectral approximation can be checked by computing the average normalised residual defined as $\dfrac{1}{N}\sum_{i=1}^{N}\dfrac{|\mathcal{P}(f_i)-\tilde{\mathcal{P}}(f_i)|}{\mathcal{P}(f_i)}$ for multiple power spectral shapes approximated by $J$ basis functions over a grid of frequencies given by $f_\mathrm{stop}/f_\mathrm{start}=T/\Delta t/2$. In \cref{fig:accuracy}, we show the accuracy of the approximation as a function of $J$, $f_\mathrm{stop}/f_\mathrm{start}$ and the model for $\psi_4$ and $\psi_6$. We see that overall when choosing $J$ between $20$ and $30$ the approximation has a good accuracy of about $1\%$.

\section{Posterior distributions}
\label{apdx:posteriors}

\begin{figure*}
    \centering
    \includegraphics[width=1\linewidth]{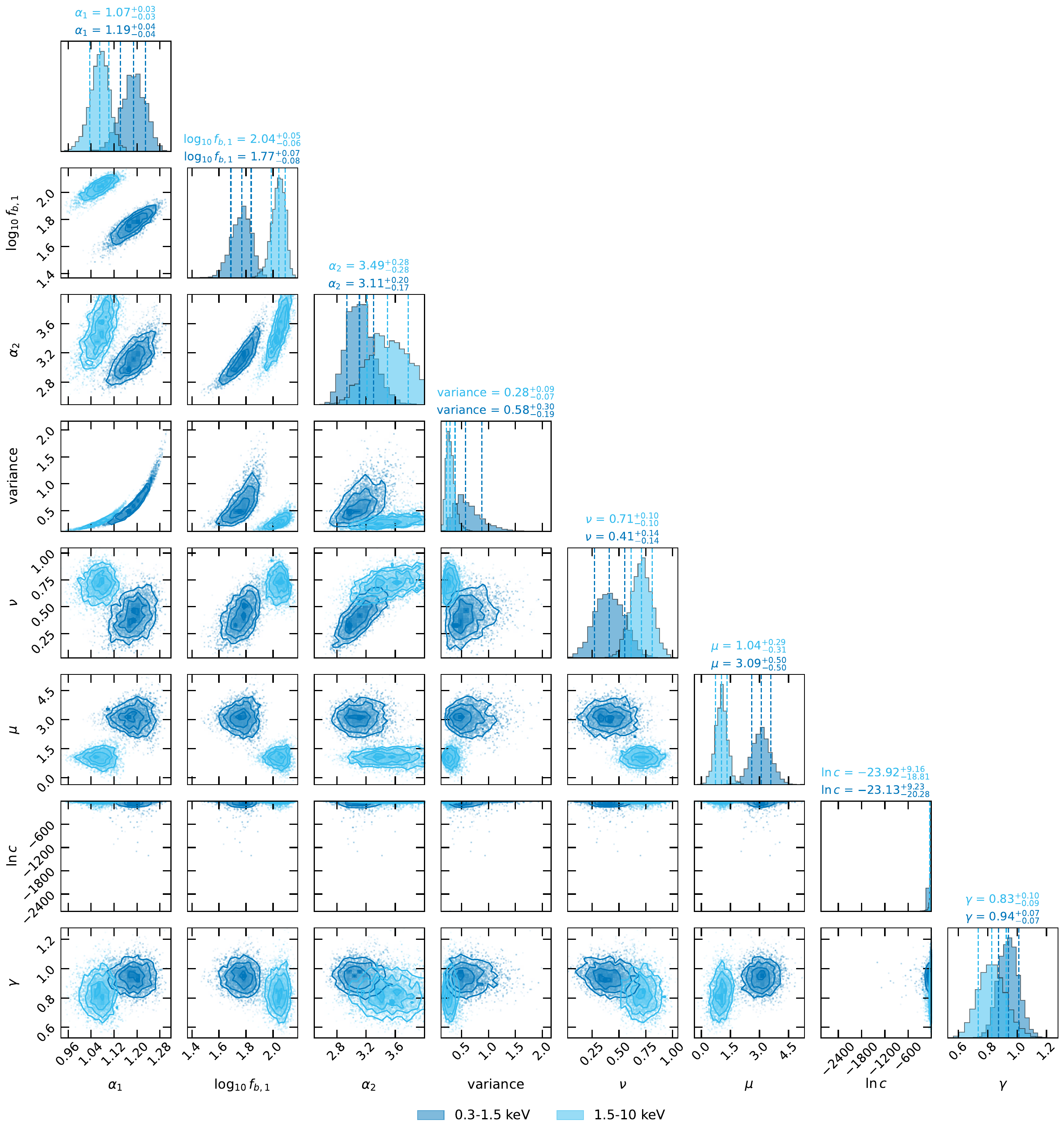}
    \caption{Posterior samples for the single bending power-law model in the soft and hard energy bands for the \textit{XMM--Newton} and \textit{Swift} observations of Ark~564.}
    \label{fig:posteriors_single}
\end{figure*}

\begin{figure*}
    \centering
    \includegraphics[width=1\linewidth]{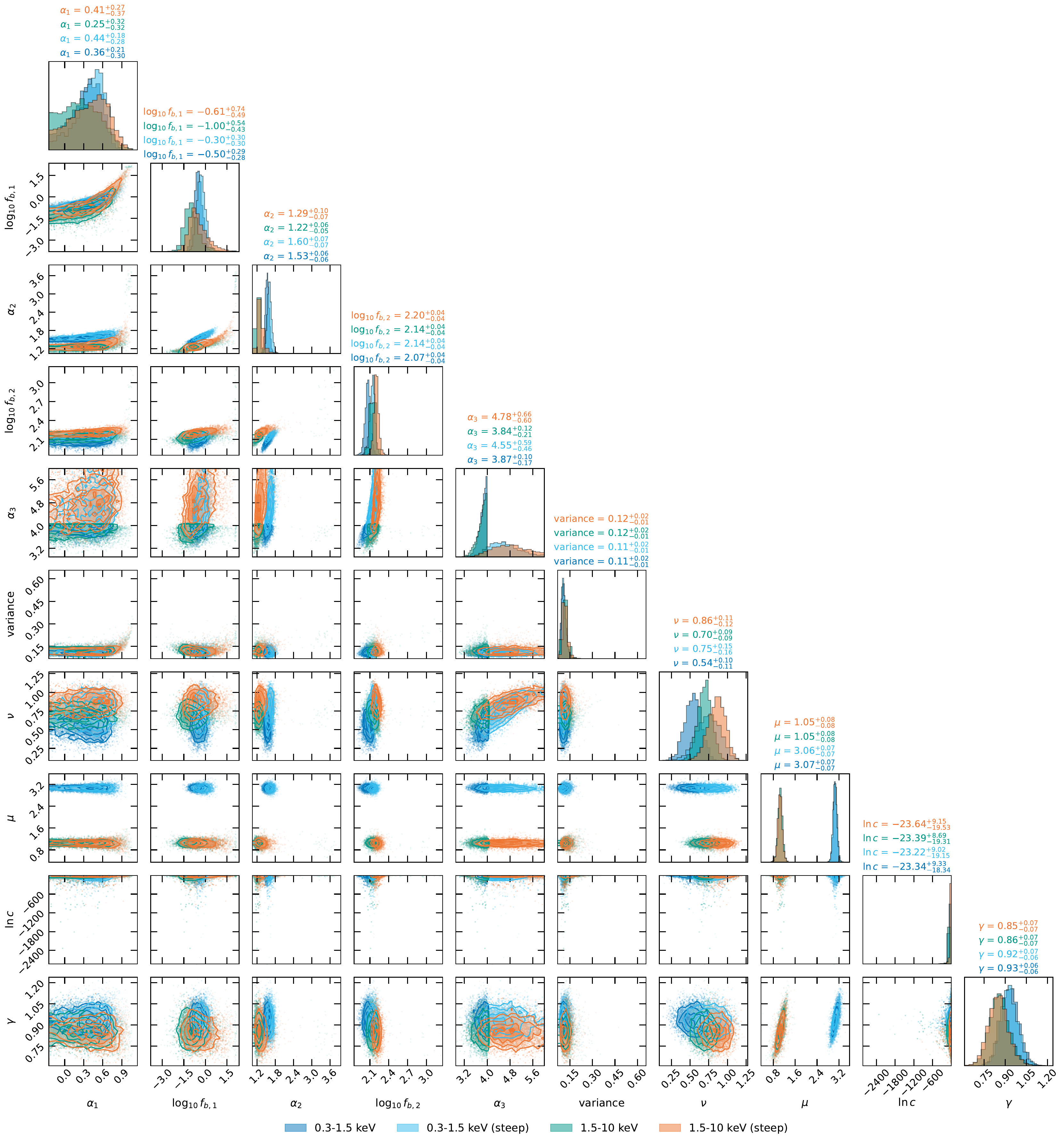}
    \caption{Posterior samples for the double bending power-law model in the soft and hard energy bands for the \textit{XMM--Newton} and \textit{Swift} observations of Ark~564.}
    \label{fig:posteriors_double}
\end{figure*}



\bsp	
\label{lastpage}
\end{document}